\documentclass[pdflatex,sn-mathphys-num]{sn-jnl}

\usepackage{graphicx}%
\usepackage{multirow}%
\usepackage{amsmath,amssymb,amsfonts}%
\usepackage{amsthm}%
\usepackage{mathrsfs}%
\usepackage[title]{appendix}%
\usepackage{subcaption}
\usepackage{xcolor}%
\usepackage{textcomp}%
\usepackage{manyfoot}%
\usepackage{booktabs}%
\usepackage{algorithm}%
\usepackage{algorithmicx}%
\usepackage{algpseudocode}%
\usepackage{listings}%
\usepackage{wrapfig}
\usepackage[capitalise]{cleveref}
\usepackage[most]{tcolorbox}
\tcbset{
    frame code={},
    center title,
    left skip=0pt,
    left=0pt,
    leftrule=-5pt,
    rightrule=-5pt,
    right=0pt,
    top=0pt,
    bottom=0pt,
    colback=black!4,
    width=1\textwidth,
    boxsep=5pt,
    arc=0pt,outer arc=0pt,
    }
\graphicspath{{figures/}}

\usepackage{thm-restate}
\theoremstyle{definition}  

\long\def\ks#1{\textcolor{black}{#1}}
\long\def\yb#1{\textcolor{black}{#1}}
\long\def\ksr#1{\textcolor{black}{#1}}
\long\def\ybr#1{\textcolor{black}{#1}}

\newcommand {\thmbox}[2]{\begin{tcolorbox}\begin{#1} #2 \end{#1}\end{tcolorbox}}

\begin{document}

\title{The Hive Mind is a Single Reinforcement Learning Agent}  


\author*[1,2]{\fnm{Karthik} \sur{Soma}}\email{karthik.soma@etud.polymtl.ca}
\equalcont{These authors contributed equally to this work.}
\author[1,2]{\fnm{Yann} \sur{Bouteiller}}
\equalcont{These authors contributed equally to this work.}
\author[3]{\fnm{Heiko} \sur{Hamann}}
\author[1,2]{\fnm{Giovanni} \sur{Beltrame}} 
\affil[1]{
\orgdiv{MIST Lab, Department of Computer and Software Engineering},
\orgname{Polytechnique Montréal},
\city{Montréal},
}
\affil[2]{
\orgname{Mila - Quebec AI Institute}
}
\affil[3]{
\orgdiv{Department of Computer and Information Science},
\orgname{University of Konstanz},
\city{Konstanz},
\country{Germany}
}


\abstract{

Decision-making is an essential attribute of any intelligent agent or group.
Natural systems are known to converge to effective strategies through at least two distinct mechanisms: collective decision-making via imitation of others, and trial-and-error by a single agent.
This paper establishes an equivalence between these two paradigms.
We show that the emergent distributed cognition (sometimes referred to as the \textit{hive mind}) arising from individuals
following simple, local imitation-based rules is that of a single online reinforcement learning (RL) agent
\ybr{aggregating action-value samples from}
many parallel \ybr{instances of a multi-armed bandit}.
\yb{More specifically, \ybr{we show that,} in the
\ybr{blindly}
imitative
\emph{weighted voter} model of
honey bees' waggle dance,}
the update rule through which this macro-agent learns is an RL
algorithm that we coin \textit{Maynard-Cross Learning}.
Our analysis implies that a group of
\ybr{individuals following simple}
\yb{imitative}
\ybr{strategies}
can be equivalent to a more complex
\ybr{learning}
entity, substantiating the idea that group-level intelligence may explain how
\ybr{seemingly irrational}
individual behaviors are selected in nature.
Beyond biology, the
framework offers new tools for analyzing economic and social systems where
individuals imitate successful strategies, effectively participating in a
collective learning process.
Our findings \ks{may further} inform the
design of scalable RL-inspired collective systems in artificial domains.

}




\maketitle

\section{Introduction}
\label{sec:intro}

\begin{figure}[htbp]
  \centering
  \def\svgwidth{\linewidth}
  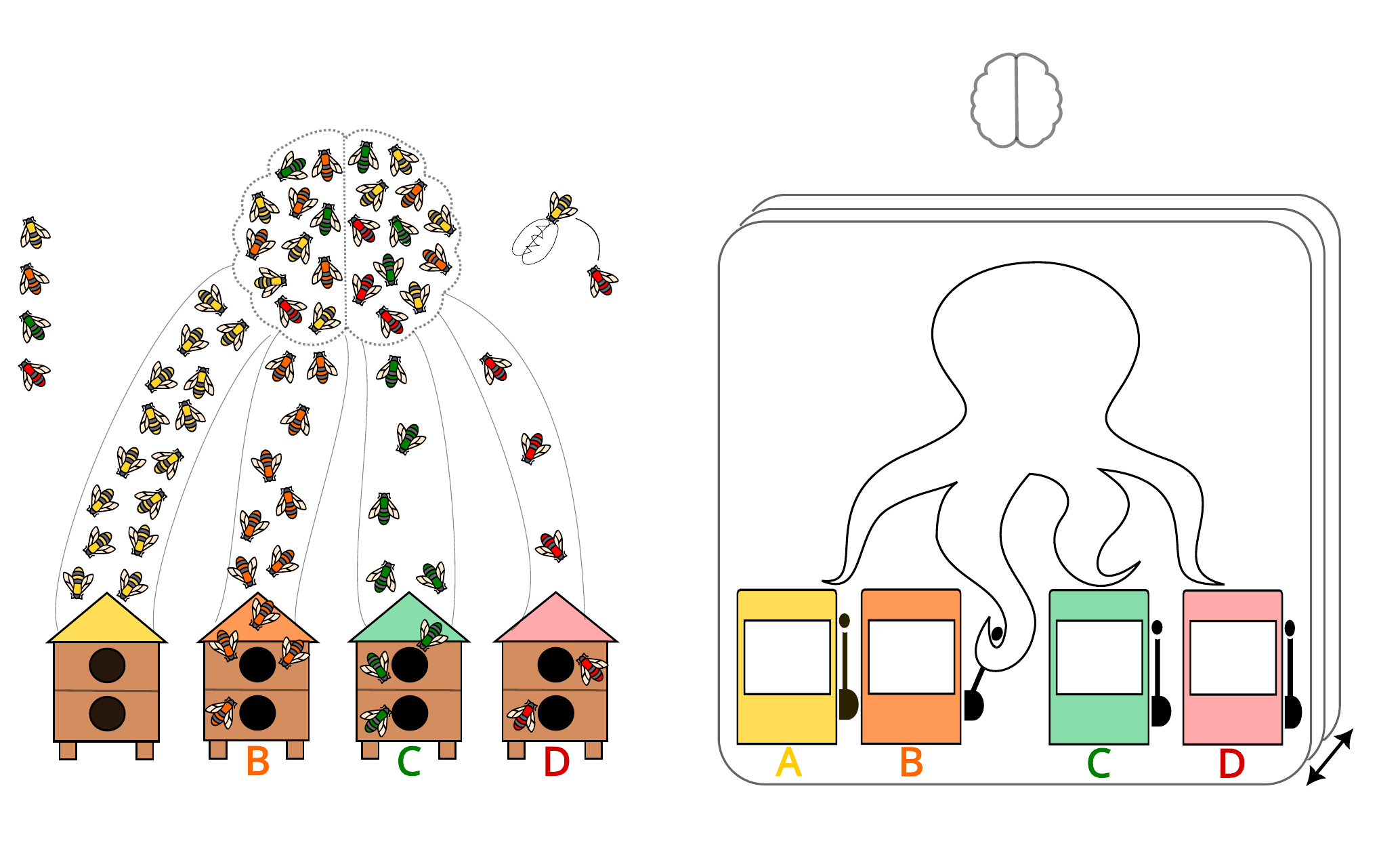
  \caption{The ``hive mind'' of a swarm of $N$ bees  nest-hunting among $n$ options is a single $n$-armed bandit RL agent learning from $N$ environments in parallel.}
\label{fig:intro}
\end{figure}


In nature, at least two distinct mechanisms allow single agents or groups of individuals\footnote{To avoid confusion, we use ``agent'' in the context of RL and ``individual'' in the context of a population.} to converge towards making optimal decisions. 
The first of these mechanisms adopts an individualistic approach, whereby the agent \textit{learns} to make optimal decisions through trial and error. This paradigm is known as Reinforcement Learning (RL),
where the agent learns a policy that maximizes the rewards it receives upon interacting with its environment~\cite{sutton2018reinforcement}.
Numerous studies in computational neuroscience have established links between learning processes happening in the brains of living beings and the formal framework of algorithmic RL~\cite{neftci2019reinforcement, Muller2024}.
In this paper, we are specifically interested in multi-armed bandits~\cite{sutton2018reinforcement}, \yb{which are stateless environments} where an RL agent learns to make the best choice among $n$ different options (or ``arms'').
Online RL agents learn by either interacting with a single environment (a setting recently referred to as the ``streaming'' setting~\cite{elsayed2024streamingdeepreinforcementlearning,vasan2024deeppolicygradientmethods}), or, when possible, by collecting samples in parallel from multiple copies of the same environment simultaneously (the ``parallel'' setting \cite{mnih2016asynchronous}).
To stabilize RL, the learning rate and the number of parallel environments are crucial in the streaming and parallel cases, respectively. Among the many learning algorithms designed to solve multi-armed bandits (Upper-Confidence-Bound~\cite{auerUCB1}, $\varepsilon$-greedy~\cite{sutton2018reinforcement}, Gradient Bandit~\cite{Williams92}, etc.),
\yb{one of the most fundamental is}
the Cross Learning (CL)~\cite{cross1973} update rule (named after the economist John G. Cross~\cite{bloembergen2015evolutionary}).

The second mechanism takes a collective approach, where individuals mimic other individuals in the group to make decisions.
Typically, in this process, 
\yb{a sufficient number of} individuals in a population
\yb{need} to agree on the best decision from a set of alternatives.
This problem is commonly referred to as best-of-n collective decision-making (CDM).
CDM arises in various domains, such as honey bee colonies~\cite{Reina15,bose2017}, human societies~\cite{Jackson10}, and robot swarms~\cite{valentini2016collective,valentini2014self}. \emph{Imitation of success}~\cite{sandholm2010population}, from Evolutionary Game Theory (EGT), is a solution to CDM where individuals mimic randomly chosen neighbors based on how successful those neighbors' decisions have been.
Another solution to CDM is the \emph{weighted voter model}~\cite{valentini2014self,Reina_2024}, which 
\yb{models \ybr{passive} imitation of \ybr{actively} advertised strategies.}
\ybr{This model is inspired by}
the nest-hunting behavior of honey bees~\yb{\cite{scott_99, seeley1978nest, seeley2010honeybee}}.
During nest-hunting, \yb{individual} honey bees
are known to
adopt
\yb{several behavioral processes~\cite{Seeley12}, the most salient being the ``waggle dance'' recruitment process}
represented in \cref{fig:intro} (left-hand half):
after scouting one of $n$ potential nesting areas, bees come back to the initial location of the swarm and perform a waggle dance that describes the coordinates of the option they have explored~\yb{\cite{seeley1978nest, scott_99}}.
This dance is performed
\yb{for a duration}
that is proportional to the estimated quality of the explored area.
Other bees go scout the area corresponding to the first dance they witness, and this process repeats until the colony reaches a quorum~\cite{seeley2004quorum},
at which point the entire swarm takes off and leaves for the \ks{chosen} site
(note that this is a simplified description of \yb{real} bees' behavior \yb{during nest-site selection; we discuss the biological limitations of this model in \cref{sec:limitation_bio} and we refer the reader to}~\ks{\cite{seeley2004group,seeley2004quorum,passino2006modeling,Passino2008,seeley2010honeybee,Seeley12}} for \yb{more} details).
We refer to \ybr{the distributed cognition emerging from} this behavior as that of a ``hive mind''~\ybr{\cite{jones2015hive}}, motivated by two factors:
(1)
\yb{biologists have already described insect colonies as agentic superorganisms~\cite{detrain1999information, baddeley2019optimal, Hunt20, hunt2020bayesian} with similarities to the neuron-based brains of vertebrates~\cite{Passino2008, marshall09, reina2018psychophysical},}
and (2) we formally show that,
\ybr{when individuals are modeled as passive imitators and active promoters according to the weighted voter model, the population}
collectively learns as if it were one single RL entity.

In this paper, we
\yb{develop a theoretical framework bridging local imitation and agentic trial-and-error}
via different variants of the Replicator Dynamic (RD)~\cite{sandholm08}.
In particular, we show that
\yb{the weighted voter model}
amounts to a single-agent, swarm-level RL process, as illustrated in \cref{fig:intro}.
To the best of our knowledge, this is the first work to formally establish this equivalence.
Specifically, our contributions are:
\begin{itemize}
    \item  \yb{As a preliminary observation,} we remark that a population of non-learning individuals following the \emph{imitation of success model} can equivalently be seen as a single agent following the \emph{Cross Learning} RL update \ybr{in a multi-armed bandit}.
    \item \yb{As our main result,} we show that the \emph{weighted voter model},
    \yb{a simplified model of the imitative behavior observed in honey bees during nest-hunting,}
    also aggregates \ks{into} a single-agent RL algorithm \ybr{solving a multi-armed bandit} at the macro-organism level.
    We coin \yb{this algorithm} \emph{Maynard-Cross Learning}.
    \item We extend our theoretical analysis with further results from simulation.
\end{itemize}
The first two contributions, which detail how social imitation models aggregate into a single reinforcement learning agent, provide a formal description of diverse instances of collective intelligence~\yb{\cite{vega1997evolution, Reina_2017}}.

\section{Background}

\subsection{Multi-armed bandits and Cross Learning (J.G. Cross)}\label{prelim:cl}

Multi-armed bandits (\cref{fig:intro}, right-hand half) are the simplest type of environments encountered in RL literature.
They consist of a discrete set of available actions, called ``arms", among which the agent has to find the most rewarding.
In the $n$-armed bandits considered in this paper, pulling an arm $a \in \{1, \dots, n \}$ returns a real-valued reward $r_a \in [ 0, 1 ]$ sampled from a hidden distribution~$r$.
The objective for an RL agent playing a multi-armed bandit is to learn a policy, denoted by the probability vector \ks{$\pi=(\pi_1,\dots,\pi_n)$}, that maximizes the rewards obtained upon pulling the arms.
Different optimization strategies exist to find such policies, one of the oldest being Cross Learning~\cite{cross1973}:

\thmbox{definition}{
Let $k$ be an action and $r_k$ a corresponding reward sample ($r_k \sim r(k, \yb{\pi})$).
Let $\pi_a$ denote the $a^{\text{th}}$ component of $\pi$.
Cross Learning (CL) updates the policy $\pi$ as:
\begin{equation}\label{eq:cl}
	\forall a, \pi_a \leftarrow  \pi_a + r_k \times\begin{cases}
		1 - \pi_a &  \text{if}\ a=k\\
		- \pi_a &  \text{otherwise}
	\end{cases}
\end{equation}
}

\noindent
For convenience, when sampling reward $r_k$ from action $k$, we denote the expected policy update on action $a$'s probability $\pi_a$ as:
\begin{equation} \label{eq:ecl}
    d\pi_a(k) = \mathbb{E}_{r_k \sim r(k, \yb{\pi})} [r_k] \times \begin{cases}
		 1 - \pi_a &  \text{if}\ a=k\\
		 - \pi_a &  \text{otherwise}
	\end{cases}
\end{equation}

\noindent
In CL, every reward $r_k$ sampled by applying an action $k$ directly affects the probabilities of all actions.
CL is
the ancestor of
the Gradient Bandit~\cite{sutton2018reinforcement} algorithm, which performs a similar update at the parameter level (called ``preferences'') of a parametric policy rather than directly updating the probability vector.
\yb{Notice that we consider an optional dependence of the $r$ distribution on $\pi$, to capture settings where rewards depend on the agent itself (as in, e.g., self-play~\cite{silver2017mastering}) in the interest of \cref{sec:policy_dependence}.}



\subsection{Evolutionary Game Theory}\label{prelim:EGT}
Evolutionary Game Theory (EGT) is the study of population games \ks{\cite{sandholm2010population, Hofbauer_Sigmund_1998, weibull95}}.
A population $\mathcal{P}$ is made of a large number of individuals, where any individual $i$ is associated with a \emph{type}, denoted by $T_i \in \{1, \dots, n \}$.
The \emph{population vector} $\pi = (\pi_1,\dots,\pi_n)$ represents the fraction of individuals in each type ($\sum_i \pi_i = 1$). Individuals are repeatedly paired at random to play a game, each receiving a separate payoff.
Individuals adapt their type based on their payoff according to a revision protocol.

\thmbox{remark}
{
\label{re:equivalence}
Population-policy equivalence.
For the argument of our paper, it is interesting to interchangeably define $\pi = (\pi_1,\dots,\pi_n)$ both as a multi-armed bandit RL policy and as a population vector \cite{bloembergen2015evolutionary}.
This is possible because in both cases the vector $\pi$ is constrained to the probability simplex.
Further, note that uniformly sampling an individual of type $a$ from the population $\mathcal{P}$ (represented by the population vector~$\pi$) is equivalent to sampling an action $a$ from the policy $\pi$.
}

\subsubsection{\yb{Social imitation framework}} \label{sec:modeling}
\yb{We study the group-level impact of local imitation through two revision protocols: \emph{imitation of success} and \emph{weighted voters}.
Traditionally, the literature modeling imitation in large populations implicitly makes the two following assumptions~\cite{weibull95, schlag1998imitate, sandholm2010population}:
\begin{itemize}
    \item The imitation process has access to the expected quality (also called ``fitness'')  $q^\pi_k$ of any type $k$ in the current population described by $\pi$,
    \item Imitation is decoupled from evaluation, i.e., an individual of type $a$ imitates an individual of type $b$ based on estimates of $q^\pi_a$ and $q^\pi_b$
    (see \cref{app:decoupling}).
\end{itemize}
We relax the first assumption: in our work, any individual of type $k$ estimates its current quality $q^\pi_k := \mathbb{E}[r_k]$ by collecting a stochastic sample $r_k \sim r(k, \pi)$.
Notice the dependence on $\pi$.
In general population games, the payoff structure is non-stationary as it depends on the current state of the population, which evolves over time.
In the interest of simplicity, we will first focus our discussion around ``games against nature'' (i.e., stationary multi-armed bandits) where $r_k \sim r(k)$ instead solely depends on $k$.
We will then discuss general population games in \cref{sec:policy_dependence}.
}

\subsubsection{Imitation of success and the Taylor Replicator Dynamic}\label{prelim:success}
In evolutionary dynamics, an important revision protocol is \emph{imitation of success}:

\thmbox{definition}
{
In the ``imitation of success'' revision model $R_{\text{success}}$, any individual $i \in \mathcal{P}$ of type $T_i=a$ executes the following process:
\begin{itemize}
    \item \yb{$i$ estimates the current quality of its type $r_a \sim r(a, \pi)$, where $0 \leq r_a \leq 1$.}
	\item $i$ samples a random individual $j \sim \mathcal{U}(\mathcal{P})$ to \emph{imitate}. Let $T_{j}$ be $b$ \yb{and $r_b \sim r(b, \pi)$}.
    \item $i$ switches from type $a$ to type $b$ with probability
    $r_b$. 
\end{itemize}
}

\noindent
One can easily see why this rule is called ``imitation of success'': $i$ imitates $j$ based on $j$'s payoff.
Imitation of success is typically used to model replication of the fittest in evolution for biology~\cite{sandholm2010population}, or certain human behaviors in economics~\cite{apesteguia2007imitation}.
When aggregated across the population, this revision model yields a famous evolutionary dynamic known as the \emph{Taylor Replicator Dynamic}~\cite{taylor78,sandholm08} (TRD) (see \Cref{lem:PG}):

\begin{equation}\label{eq:trd}
    \dot \pi_a = \pi_a (q^\pi_a - v^\pi),
\end{equation}
where
$\dot \pi_a$ is the derivative of the $a$-th component of the population vector,
$q^\pi_a := \mathbb{E}[r_a]$ is the expected payoff of the type $a$ against the current population,
and
$v^\pi := \sum_{b}\pi_b\mathbb{E}[r_b]$ is the current average payoff of the entire population.

\subsubsection{Weighted Voter Model and the Maynard Replicator Dynamic}\label{prelim:wvoter}  

The \emph{weighted voter} model~\cite{scott_99, valentini2014self, Reina_2024} instead
\yb{models \ybr{passive imitation of actively advertised strategies, similar to} the waggle dance recruitment process observed in honey bees during nest-hunting (see \cref{sec:intro}).
\ybr{We will thus loosely refer to weighted voters as ``bees'' for the purpose of illustrating our theoretical analysis.
The reader should however keep in mind that}
this model is only a partial description of the behavior of real bees, which comprises other, non-imitative processes such as spontaneous commitment and cross-inhibition~\cite{Seeley12} \ybr{(see \cref{sec:limitation_bio})}.
We leave derivations from more exhaustive biological models of bee interactions for future work~\cite{Reina_2017}.}

\thmbox{definition}
{
\label{def:weighted}
In the ``weighted voter'' revision model $R_{\text{wvoter}}$, any bee $i \in \mathcal{P}$ of type $T_i=a$ (where $a$ corresponds to a nest-site option) executes the following process:
\begin{itemize}
    \item $i$ estimates the quality of its current type $r_a \sim r(a, \yb{\pi})$, where $0 \leq r_a \leq 1$.
    \item After obtaining $r_a$, $i$ locally broadcasts its type at a frequency proportional to $r_a$.
    \item $i$ switches its type to the first type $b$ that it perceives from its neighborhood.
    Assuming all individuals are well-mixed in the population~\cite{martin06}, the corresponding expected probability of $i$ switching to type $b$ is the proportion of votes cast for $b$ within its neighborhood: $$P^{(i)}_\text{neighborhood}(b\leftarrow a) = \frac{N^{(i)}_b\mathbb{E}[r_b]}{\sum_l N^{(i)}_l\mathbb{E}[r_l]}$$ where $N^{(i)}_k$ is the number of individuals of type $k$ in the neighborhood of $i$.
\end{itemize}
}

\noindent
\yb{The model is discretized into time-steps, and the statistical effect of the duration of a waggle dance is abstracted as a broadcasting frequency.}
Note that, in this model, honey bees do not need to directly observe the quality estimate of other scout bees, but only their type.
In \cref{sec:bees_mrd}, we show that the weighted voter revision model aggregates \ks{into} a variant of the TRD, called the \emph{Maynard-Smith Replicator Dynamic}~\cite{smith82} (MRD):  
\begin{equation}\label{eq:mrd}
    \dot \pi_a = \frac{\pi_a}{v^\pi} (q^\pi_a - v^\pi)
\end{equation}


\section{Methodology} \label{sec:theory}

In \cref{sec:imitation_cl}, we remark that, considered together, previous results from the literature yield an interesting insight binding $R_\text{success}$ and reinforcement learning: a large population following the ``imitation of success'' revision model can equivalently be considered as a single \ks{multi-armed bandit} RL agent.
While this is a simple consequence of previously known results, we could not find this insight formulated in the literature, and thus we formalize it as \cref{prop:taylor}.
Then, in \cref{sec:bees_mrd}, we prove that the ``weighted voter'' revision model also aggregates \ks{into} a \ks{multi-armed bandit} RL algorithm.
In other words, our analysis indicates that, at least in the
\yb{simplified model of nest-hunting described by $R_\text{wvoter}$,}
a swarm of \ks{honey bees} collectively acts as a single RL entity.
We formalize this result as \cref{prop:maynard}.

\subsection{Imitation of Success and Cross Learning}
\label{sec:imitation_cl}

Evolutionary Game Theorists have long been interested in the ``imitation of success'' revision protocol, as it models replication of the fittest in the evolutionary setting.
In this literature, it is famously known that a population of individuals following $R_{\text{success}}$ aggregates \ks{into} the Taylor Replicator Dynamic (see for instance \cite{sandholm08}).
\yb{Our proof in \cref{sec:apx_proof} subtly differs from the one found in \cite{sandholm08}, as we do not assume that individuals imitating each other have access to ground-truth qualities.}

\begin{restatable}{lemma}{PG}
\label{lem:PG}
An infinite population of individuals adopting $R_{\text{success}}$ follows the TRD:
\begin{equation}
d\pi_a = \pi_a (q^\pi_a - v^\pi),
\end{equation}
where $\pi_a$ is the proportion of type $a$ in the population, $q^\pi_a$ is the expected payoff (also called ``fitness'') of type $a$
\yb{given the state of}
the population, and $v^\pi$ is the average fitness of the
population.
\end{restatable}


\noindent
Our choice of notation in \cref{lem:PG} is reminiscent of the RL literature and may seem unusual from an EGT perspective.
In fact, this choice is motivated by another, lesser-known result from~\cite{borgers1997learning}, who showed that the Cross Learning RL algorithm performs updates that are also similar to the TRD (proof in \cref{sec:apx_proof}):
\begin{restatable}{lemma}{RL}
\label{lem:RL}
In expectation, an RL agent learning via the CL update rule follows:
\begin{equation}\label{eq:lem1}
    \mathbb{E}[ d\pi_a] = \pi_a (q^\pi_a - v^\pi),
\end{equation}
where $q^\pi_a$ is the action-value of $a$, and $v^\pi$ is the value of policy $\pi$.
\end{restatable}

\noindent
With this notation, it is straightforward to combine \cref{lem:PG,lem:RL}, provided a sensible duality exists between the identical terms.
\ks{As described in} \cref{re:equivalence}, a duality \ks{exists} between the population vector of \cref{lem:PG} and the policy of \cref{lem:RL}, $\pi$.
In \cref{lem:PG}, the fitness $q^\pi_a$ is the payoff that an individual of type $a$ can expect on average when encountering a random individual from the population $\pi$.
In \cref{lem:RL}, the dual of this individual of type $a$ is an action $a$ sampled from the policy $\pi$, whose expected reward is the action-value $q^\pi_a$.
Similarly, the dual of the population fitness $v^\pi$ is the policy-value.
In other words, under $R_{\text{success}}$, individuals sampled from the population $\pi$ can equivalently be regarded as action samples from an RL macro-agent, whose rewards are the individuals' payoffs.
This agent maximizes the average payoff of the group via Exact Cross Learning (we call ``exact'' the RL algorithm that directly applies expected updates instead of updates computed from sample estimates):
\thmbox{prop}
{
\label{prop:taylor} \label{lem:taylor}
An infinite population of individuals following $R_{\text{success}}$ can equivalently be seen as an RL agent following Exact Cross Learning, i.e.,
\begin{equation}
    d^\text{success}\pi_a = \mathbb{E}[ d^\text{CL}\pi_a]\;,
\end{equation}
where $\pi$ is both a population vector and a vector of action-probabilities, $d^\text{success}\pi$ is the single-step change in the population vector $\pi$ under the ``imitation of success'' revision model
, and $d^\text{CL}\pi$ is an update performed by CL on the policy $\pi$.
}
\begin{proof}
    Direct consequence of \cref{lem:PG,lem:RL}.
\end{proof}

\noindent
\ybr{\cref{prop:taylor} motivates the concept of ``hive mind'' that we define as follows:}
\thmbox{definition}
{
\label{def:hive_mind}
\ybr{
Let a population $\mathcal{P}$ be defined by its population vector $\pi$. We refer to the corresponding policy $\pi$ and its learning dynamic $d\pi$ as the ``hive mind'' of $\mathcal{P}$.
}
}

\noindent
\ybr{\cref{prop:taylor} shows that the ``hive mind'' of a population of individuals following $R_{\text{success}}$ implements Exact Cross Learning.}
Intuitively, this macroscopic RL process can appear to be a mere by-product of individuals in the population imitating others to optimize for their own success.
But let us now turn our attention to organisms that may not even have a notion of individual success to optimize: honey bees.


\subsection{Weighted Voters and Maynard-Cross Learning}
\label{sec:bees_mrd}

We now consider a large population of $N \gg 1$
\yb{purely imitative ``honey bees''}
with average local neighborhood size $M \gg 1$, seeking agreement on which nesting site to select by applying~$R_{\text{wvoter}}$.
\yb{We further assume that, after evaluating their type, scouts land and walk randomly in the bee cluster, yielding a \emph{well-mixed} population state~\cite{martin06, valentini2016collective}.}
We show that, although $R_{\text{wvoter}}$ is a blind imitation protocol where individuals simply imitate the first type they encounter, a swarm following $R_{\text{wvoter}}$ aggregates \ks{into} an RL agent at the macro-organism level.
In the~$R_{\text{wvoter}}$ model, individual bees are \ybr{modeled as} blind imitators, \ybr{but also as} active promoters.
Each bee $i$ of type $T_i = k$ and payoff sample $r^{(i)}_k \sim r(k, \yb{\pi})$ has a tangible stochastic influence on the local expected inflow of other bees that it rallies to its own type $k$ within its local neighborhood $\mathcal{N}_i$:
\begin{equation}
    P^{(i)}_\text{neighborhood}(k \leftarrow \cdot) = \frac{r^{(i)}_k}{\sum\limits_{j \in \mathcal{N}_i} r^{(j)}},
\end{equation}
where $r^{(i)}_k$ represents $i$'s broadcasting frequency, and $\sum_{j \in \mathcal{N}_i} r^{(j)}$ represents the total broadcasting frequency of $i$'s local neighborhood.
The expected outflow attributable to $i$ on the entire swarm from any type $a \neq k$ to the type $k$ is thus:
\begin{align}
    P^{(i)}(k \leftarrow a) &= \frac{N^{(i)}_a}{N} P^{(i)}_\text{neighborhood}(k \leftarrow \cdot) \notag \\ &= \frac{N^{(i)}_a}{N} \frac{r^{(i)}_k}{\sum\limits_{j \in \mathcal{N}_i} r^{(j)}},
\end{align}
where $N^{(i)}_a$ is the number of type-$a$ bees within $i$'s neighborhood.
For simplicity, we assume that $M$ (and thus also $N$) is large.
Assuming that all individuals are well-mixed in the population, it follows that $\frac{N^{(i)}_a}{M} = \pi_a$:
\begin{align} 
P^{(i)}(k \leftarrow a) &= \frac{M \pi_a}{N} \frac{r^{(i)}_k}{\sum\limits_{j \in \mathcal{N}_i} r^{(j)}} \notag\\
&= \frac{\pi_a}{N} \frac{r^{(i)}_k}{\frac{1}{M}\sum\limits_{j \in \mathcal{N}_i} r^{(j)}} \notag\\
&= \alpha \frac{r^{(i)}_k}{v^\pi} \pi_a,\label{eq:mcl1}
\end{align}
where $\alpha := \frac{1}{N}$ and $\frac{1}{M}\sum\limits_{j \in \mathcal{N}_i} r^{(j)}$ is the average payoff of the individuals in 
$i$'s neighborhood.  In a well-mixed population, this is also the average population payoff 
$v_{\pi}$. 
Summing over all types except $k$ (whose outflow to $k$ is 0), we obtain the total inflow into type $k$ attributable to $i$:
\begin{align} 
\sum_{a \neq k} P^{(i)}(k \leftarrow a) &= \sum_{a \neq k} \alpha \frac{r^{(i)}_k}{v^\pi} \pi_a \notag\\ &= \alpha \frac{r^{(i)}_k}{v^\pi}\sum_{a \neq k}  \pi_a \notag\\
&= \alpha \frac{r^{(i)}_k}{v^\pi} (1 - \pi_k)\;.\label{eq:mcl2}
\end{align}

\noindent
Given a bee $i$ of type $k$, \cref{eq:mcl1} describes the outflow that its
\ybr{weighted vote}
induces within the swarm from any  type $a$ to the type $k$, while \cref{eq:mcl2} describes the corresponding inflow into type $k$ from all types.
In other words, the bee $i$ can be seen as an action-reward sample whose influence on the population vector $\pi$ is described by an RL update rule that we coin \emph{$\alpha$-Maynard-Cross Learning} ($\alpha$-MCL):

\thmbox{definition}
{
\label{def:mcl_alpha}
Let $k$ be an action and $r_k \sim r(k, \yb{\pi})$ a corresponding reward sample.
$\alpha$-MCL updates the policy $\pi$ as:
\begin{equation} \label{eq:mcl_alpha}
	\forall a, \pi_a \leftarrow  \pi_a + \alpha \frac{r_k}{v^\pi}\begin{cases}
		1 - \pi_a &  \text{if}\ a=k\\
		- \pi_a &  \text{otherwise}
	\end{cases}
\end{equation}
where $v^\pi$ is the current value of policy $\pi$.
}

\noindent
Here, the meaning of $\alpha$ is of peculiar interest.
In \cref{eq:mcl_alpha}, $\alpha$ looks very much like a learning rate, i.e., a hyperparameter that RL practitioners typically set to a small value in order to downsize the amplitude of individual policy updates.
When dealing with on-policy RL algorithms, they often evaluate the policy in several parallel copies of the same environment~\cite{makoviychuk2021isaacgymhighperformance}, average the corresponding updates into a single policy update, and downsize this update via a small learning rate.
The common rule of thumb is that the learning rate can be larger when there are enough parallel environments.
But in \cref{def:mcl_alpha}, $\alpha$ instead has a clear population-based meaning: it is the inverse of the number of bees acting in parallel in the swarm, as we are describing the influence of one single bee on the population.
Crucially, this influence is that of an RL agent, but
\yb{in the weighted voter model}
the individual bee itself is
\ybr{not modeled as}
an RL agent: it
only
\ybr{``blindly''}
imitates its peers, \ybr{irrespective of their payoffs}!
Instead,
\ybr{social imitation may evolve in nature}
because it aggregates \ks{into} an RL agent at the macro-organism level: the ``hive mind'', whose rewards are the
\ybr{parallelly sampled quality estimates.}
Summing the influence of all individual bees yields:
\begin{align} 
    d\pi_a &= \sum_{\ks{i=1}}^N \frac{r^{(i)}}{N v^\pi} \begin{cases}
		1 - \pi_a &  \text{if}\ a=k\\
		- \pi_a &  \text{otherwise}
	\end{cases}\notag\\
    &= \sum_k \frac{N_k q^\pi_k}{N v^\pi} \begin{cases}
		1 - \pi_a &  \text{if}\ a=k\\
		- \pi_a &  \text{otherwise}
	\end{cases}\notag\\
&= \frac{1}{N v^\pi}(N_a q^\pi_a (1-\pi_a) - \sum_{k \neq a} N_k q^\pi_k \pi_a)\notag\\
&= \frac{1}{N v^\pi}(N_a q^\pi_a - \sum_{k} N_k q^\pi_k \pi_a)\notag\\
&= \frac{1}{v^\pi}(\pi_a q^\pi_a - \pi_a \sum_{k} \pi_k q^\pi_k)\notag\\
&= \frac{\pi_a}{v^\pi}(q^\pi_a - v^\pi)\label{eq:summed_mcl}
\end{align}
which is the Maynard-Smith Replicator Dynamic.
Similar to \cref{lem:taylor}, we can further show that this update is the expected update performed by the 1-MCL algorithm, that we simply call \emph{Maynard-Cross Learning} (MCL) for conciseness:
\begin{align}
\mathbb{E}[d^\text{MCL}\pi_a] &= \sum_{k=1}^n \pi_k . d^\text{MCL}\pi_a(k)\notag\\
   &= \pi_a . d^\text{MCL}\pi_a(a) + \sum_{k \neq a} \pi_k . d^\text{MCL}\pi_a(k) \notag\\
    &= \pi_a \frac{\mathbb{E}[r_a]}{v^\pi}(1 - \pi_a) + \sum_{k \neq a} \pi_k \frac{\mathbb{E}[r_k]}{v^\pi}(- \pi_a)\notag\\
    &= \frac{\pi_a}{v^\pi}\Bigr[\mathbb{E}[r_a] - \pi_a \mathbb{E}[r_a] - \sum_{k \neq a} \pi_k \mathbb{E}[r_k]\Bigr]\notag\\
    &= \frac{\pi_a}{v^\pi}\Bigr[\mathbb{E}[r_a] - \sum_{k} \pi_k \mathbb{E} [r_k]\Bigr]\notag\\
    &= \frac{\pi_a}{v^\pi} (q^\pi_a - v^\pi)\label{eq:expected_mcl}
\end{align}
Thus we can write the RL update of the ``hive mind'' in a swarm
following the $R_{wvoter}$ model of local imitation:
\thmbox{prop} 
{
\label{prop:maynard}
An infinite population of individuals following $R_{\text{wvoter}}$ can equivalently be seen as an RL agent following Exact Maynard-Cross Learning, i.e.,
\begin{equation}
    d^\text{wvoter}\pi_a = \mathbb{E}[d^\text{MCL}\pi_a],
\end{equation}
where $\pi$ is both a population vector and a vector of action-probabilities, $d^\text{wvoter}\pi$ is the single-step change in the population vector $\pi$ under the weighted voter revision model, and $d^\text{MCL}\pi$ is the update performed by MCL on the policy $\pi$.
}
\begin{proof}
    Direct consequence of \cref{eq:summed_mcl,eq:expected_mcl}.
\end{proof}

\section{Simulations and final remarks}\label{sec:results}

\begin{figure}[t]
    \centering
    \begin{subfigure}[t]{0.496\textwidth}
        \includegraphics[width=\linewidth]{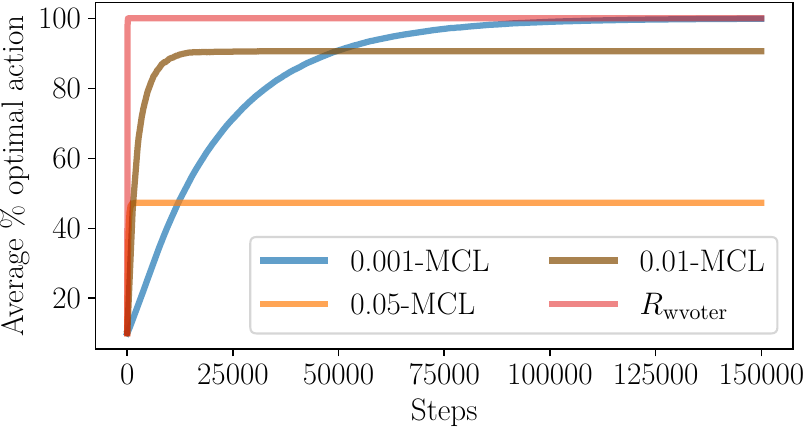}
        \subcaption{ $\alpha$-MCL for varying learning rates}
        \label{fig:mcl-alpha}
    \end{subfigure}
    \begin{subfigure}[t]{0.496\textwidth}
        \includegraphics[width=\linewidth]{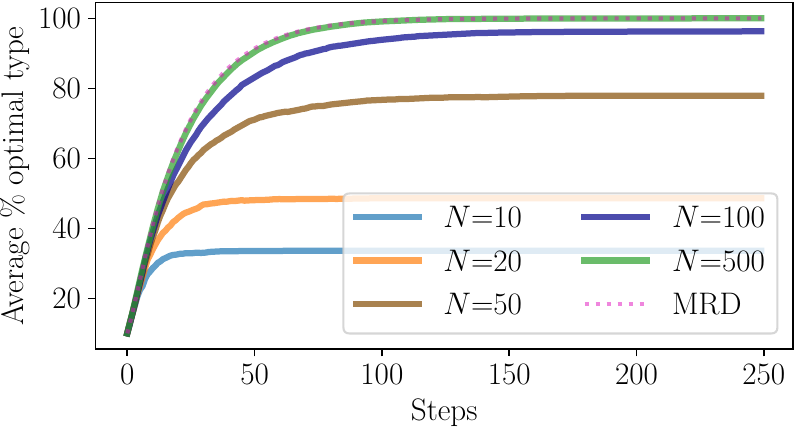}
        \subcaption{Varying swarm sizes $N$}
        \label{fig:pop_wvoter_varying}
    \end{subfigure}

    \begin{subfigure}[t]{0.496\textwidth}
        \includegraphics[width=\linewidth]{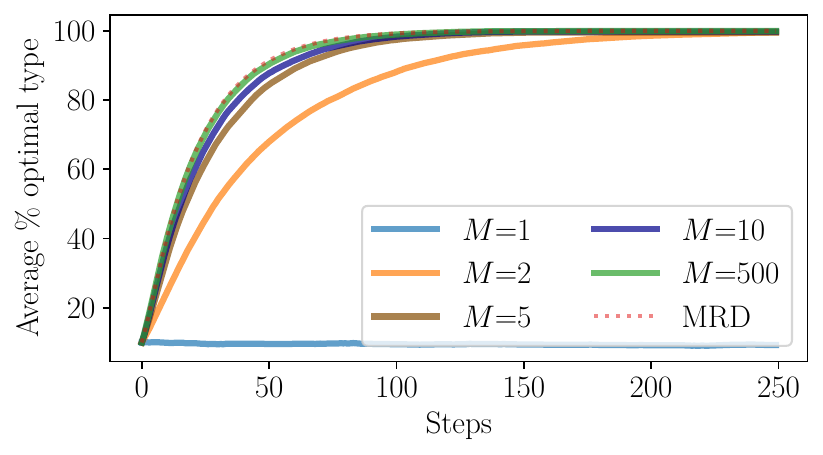}
        \subcaption{Varying neighborhood sizes for $N$=500}
        \label{fig:pop_wvoter_varying_M}
    \end{subfigure}
    \begin{subfigure}[t]{0.496\textwidth}
        \includegraphics[width=\linewidth]{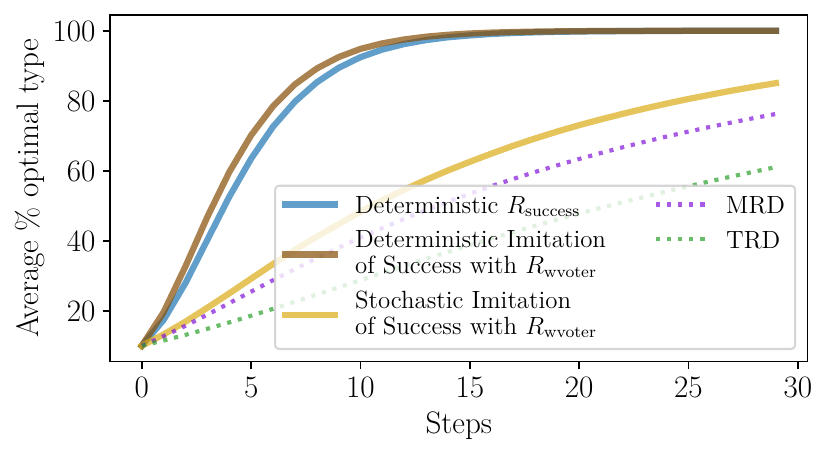}
        \subcaption{Variants of $R_{\text{success}}$ \& $R_{\text{wvoter}}$ ($N$=1000)}
        \label{fig:main_variants}
    \end{subfigure}

    \caption{\textbf{\ks{Simulations}}: (a) Swarms reach consensus more rapidly when following $R_{\text{wvoter}}$ collectively than when individuals learn via $\alpha$-MCL. (b,c) Varying swarm and neighborhood sizes show that the theoretical predictions hold under practical constraints. (d) Simple variants of $R_{\text{success}}$ and $R_{\text{wvoter}}$ can surpass $R_{\text{wvoter}}$. 
    }
    \label{fig:combined}
\end{figure}

\subsection{A single RL agent in many parallel environments}
\label{sec:lr_batching}

Researchers have already observed collective intelligence~\cite{swarmintelligence99, franks2002information}, often described as ``emergent''~\cite{de2004emergence} to describe situations where simple, \ybr{seemingly irrational} individual behaviors aggregate \ks{into} a complex and coherent behavior of the group.
\cref{prop:maynard} shows that, when the individual bees' behavior is modeled as $R_\text{wvoter}$, this ``emergent'' collective intelligence is MCL, a multi-armed bandit RL algorithm.
But the implications of \cref{prop:maynard} go even further:
\ks{a swarm of bees following $R_\text{wvoter}$ is not only a single MCL agent, but one that}
is quite efficient at what it does, because each bee is a parallel action sample of its policy.
In other words, each bee $i$ can be seen as an action sample $a^{(i)} \sim \pi$ tested against a parallel copy $E^{(i)}$ of the environment $E$, as illustrated in \cref{fig:intro}.
\cref{fig:mcl-alpha} shows that this converges much faster than if individual bees were themselves to learn via $\alpha$-MCL instead of just following $R_\text{wvoter}$.
If the MCL algorithm were of any use outside of swarm intelligence\footnote{Most likely not: more advanced algorithms exist to solve multi-armed bandits (such as UCB~\cite{auerUCB1}).}, say if a computer scientist wanted to train a robot via iterative $\alpha$-MCL, they would need a small $\alpha$ for the algorithm to converge to the optimal option.
Or, if feasible, they could afford using a larger $\alpha$ by evaluating the policy in several parallel simulations (see \cref{app:streaming_rl,app:parallel_rl}).
Swarms of honey bees naturally learn via parallel RL in as many environments as there are scouts in the swarm.
Finally, \cref{prop:taylor} implies that a similar discussion holds for the ``imitation of success'' revision protocol, which models certain human behaviors, and even natural
selection itself.

\subsection{Swarm size}



In \cref{sec:lr_batching}, we have discussed how being made of a number of individuals enables the ``hive mind'' to evaluate actions sampled from $\pi$ in a parallel fashion.
Furthermore, our analysis assumed that this number was large (which enabled us to compute flows in \cref{sec:theory}).
Yet, in practice there is a tradeoff here.
Actual honey bee swarms are known to use only a small proportion of their bees as scouts, while the large majority of the swarm remains quiescent during nest-hunting~\cite{seeley1978nest, beekman2006does}, which is surmised to be in the interest of energy-efficiency.
In a typical swarm of 10,000 bees, roughly 200 to 500 scouts are active during nest-site selection.
To analyze how the number of scouts~$N$ affects the task-performance of the MCL macro-agent, we conducted simplified simulations where swarms of varying sizes had to choose the best amongst 10 options.
Option qualities were spread between 0 and 1, while the noise in bees' estimates was modeled with a uniform perturbation of amplitude $0.2$.
\cref{fig:pop_wvoter_varying} reports the percentage of populations that converged to the optimal choice for different values of $N$, across 1000 seeds (each individual seed converged to a homogeneous choice).
In these simulations, $N=500$ is enough to closely follow the MRD and converge to the optimal option, whereas a number of scout bees that is too small often yields convergence to a sub-optimal nest site option.
However, our choice of simulation parameters is largely arbitrary: \cref{fig:pop_wvoter_varying} should be interpreted qualitatively rather than quantitatively.

\subsection{Neighborhood size}


In \cref{sec:theory}, we have assumed that the local neighborhood size $M$ was large.
This assumption was useful to analytically derive the macro-agent RL update, because it meant that local neighborhoods (\ybr{e.g.}, other scout bees that a scout may randomly walk into within its vicinity) could be considered well-mixed.
Arguably, actual neighborhood sizes are not that large in the real world.
Therefore, to complete our theoretical analysis, we performed $R_\text{wvoter}$ simulations with varying neighborhood sizes.
\cref{fig:pop_wvoter_varying_M} shows that even a small $M \geq 5$ yields a macro-agent algorithm that closely follows MCL.
With $M=1$, each scout copies an arbitrary neighbor, which yields no macro-dynamic.

\subsection{Collective power of \ks{promotion}}
The $R_\text{wvoter}$ revision protocol
has surprising advantages over imitation of success.
In \cref{sup:methods}, we show that $R_\text{wvoter}$ (MCL) generally converges faster than $R_\text{success}$ (CL), except for small population sizes.
This is in addition to the simplicity of $R_\text{wvoter}$, which only requires \ybr{individuals} to blindly mimic their peers.
\subsection{\yb{Other imitation protocols}}
In \cref{sec:theory}, we showed that the bees employing $R_{\text{wvoter}}$ can be
seen as a single online RL agent. However, a natural
question arises: is $R_{\text{wvoter}}$ the optimal collective-decision making
strategy? To answer this question, we investigate three simple variants of the
imitation of success model and the weighted voter model (more details in \cref{app:pop_details}):
\begin{itemize}
    \item Deterministic Imitation of Success: Unlike the standard imitation of success model (which switches stochastically based on partner's success), this variant deterministically adopts the
    partner's type when it has more success.
    \item Deterministic Imitation of Success with Weighted Voter Rule: This variant combines the weighted voter model's success-weighted neighbor sampling with the deterministic imitation of success switch based on comparative success.
    \item Stochastic Imitation of Success with Weighted Voter Rule: This variant differs from the previous in that it uses classic stochastic imitation of success. 
\end{itemize}
\cref{fig:main_variants} shows that, at least in this simplified environment,
all these variants converge to the optimal decision faster than
$R_{\text{wvoter}}$ or the MRD.
This
experiment suggests that there may be
better collective decision-making strategies than $R_{\text{wvoter}}$. In
particular, 
\ybr{when individuals}
perform comparisons with their
individual quality estimates
instead of
blindly imitating others from their
neighborhood,
the convergence speed
\ybr{is}
faster than the MRD.
This insight raises an \ybr{intriguing} question:
\yb{previous studies on nest-site selection in honey bees \ybr{have} found evidence that individual bees do not make direct comparisons between quality estimates \ybr{in a meaningful way}~\cite{scott_99}.
But if \ybr{not}, why don't they?}
\yb{One possible explanation \ybr{would be} that
\ybr{$R_\text{wvoter}$-style}
imitation uses minimal communication and cognition bandwidth as it}
requires no comparison operations between quality values and no
direct communication of numerical quality estimates,
\yb{contrary to all our tested}
faster-converging variants.
\yb{
\ybr{Note however that, during food foraging (rather than nest-site selection), later studies have found evidence of reinforcement happening at the level of individual bees}~\cite{biesmeijer2005use, maboudi20}, which \ybr{does} involve \ybr{performing direct comparisons between their own} quality samples.}

\subsection{\yb{Imitation in general population games}} \label{sec:policy_dependence}
\yb{
Our discussion so far has revolved around ``games against nature'', i.e., games in which individuals collect samples from their environment in a purely parallel fashion, without mutually affecting each other's payoffs.
This is reasonable for a simplified model of nest-site selection, but more complex settings require moving away from stationary bandits.
In general population games, quality estimates $r_k \sim r(k, \pi)$ do depend on both the type $k$ of the individual and the current structure of the population $\pi$.
For instance, congestion may occur in a foraging task that would reduce the qualities of overly sampled arms, or we may be interested in settings where the payoff of an individual depends on the types of opponents it encounters in the population.
In fact, as we discuss in \cref{sec:limitation_bio}, the actual nest-site selection process observed in real swarms of honey bees also depends on the population structure $\pi$.
}
\yb{
As previously noted, a dependence on $\pi$ does not affect our mathematical derivations.
Our choice of notation, $r_k \sim r(k, \pi)$, is intentionally general to capture a broad class of population-dependent quality estimation processes.
For instance, let us imagine a classical Evolutionary Game Theoretical scenario where the population is randomly split into $N/m$ groups so that each group plays an $m$-player game ($1 \leq m \leq N$).
The population is then reshuffled (see \cref{app:decoupling}) and performs an imitation step.
Propositions \ref{prop:taylor} and \ref{prop:maynard} naturally capture this setting: the corresponding macro-agent is still CL or MCL, respectively.
But the $N$ parallel payoff samples collected by this CL (resp. MCL) macro-agent are now intrinsically correlated.
This setting corresponds to self-play~\cite{silver2017mastering, berner2019dota} in $N/m$ parallel $m$-player games where the ``hive mind'' $\pi$ is evaluated against $m$ copies of itself.
Keep in mind that $\pi$ only makes sense at the macro-agent level, though. Individuals are not self-play agents: each has its own type $k \sim \pi$.
}\\
\begin{figure}[tbp]
    \centering
        \begin{subfigure}[t]{0.496\textwidth}
        \includegraphics[width=\linewidth]{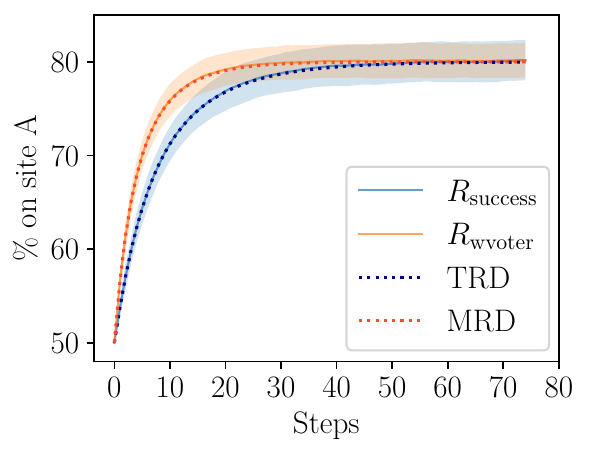}
        \subcaption{Imitation update ($N=2000$)}
    \end{subfigure}
    \begin{subfigure}[t]{0.496\textwidth}
        \includegraphics[width=\linewidth]{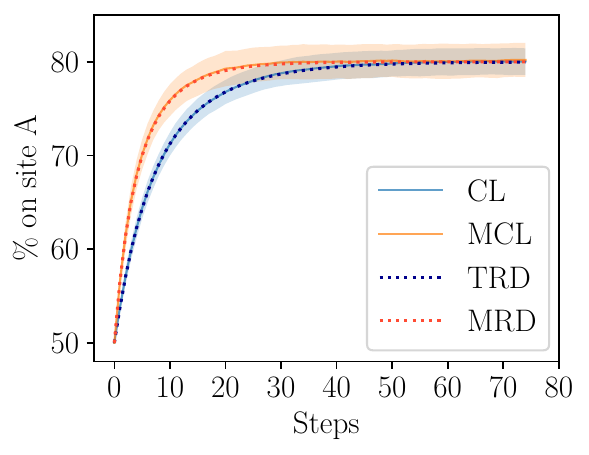}
        \subcaption{RL update ($N=2000$)}
    \end{subfigure}
    \caption{\ks{Congestion experiments in a non-stationary two-armed bandit.}}
    \label{fig:congestion}
\end{figure}

\noindent
\yb{
To illustrate how our analysis plays when quality estimates are population-dependent, we simulate the effect of congestion in a swarm of foragers that would follow either $R_{\text{wvoter}}$ or $R_{\text{success}}$.
We reduce the quality of sites proportional to the number of foragers on that site, with a congestion factor that reflects the maximum bandwidth that the site can accommodate.
This creates a non-stationary scenario where $r_k \sim r(k, \pi)$ does depend on $\pi$.
In \cref{fig:congestion}, we consider two sites with equal base qualities but differing congestion penalties, with a lower congestion penalty on site A (implementation details in \cref{app:congestion}).
We see that (1) the imitation rules $R_{\text{success}}$ (resp.\ $R_{\text{wvoter}}$) follow the TRD (resp.\ MRD), and (2) the RL update rules CL (resp.\ MCL) also follow the TRD (resp.\ MRD), assigning roughly 80\% of individuals to the site with the lowest congestion penalty.
At the equilibrium, both sites have the same effective quality, which is why we observe residual variance around 80\%.
}\\

\noindent
\ksr{
Of course,
real-world congestion penalties may be more complicated than the linear penalty model used in our simplified experiments (e.g., threshold effects, hysteresis...), and would warrant the use of embodied simulations such as~\cite{Soma2023}.
We leave this investigation
for future work.}


\subsection{\ks{Multiple populations (MARL)}}


\ks
{
\noindent
In this paper, our derivations do not capture populations of RL-enabled individuals~\cite{bouteiller2026sociodynamics, yang2018mean}, but rather populations of individuals imitating each other.
Propositions~\ref{prop:taylor} and \ref{prop:maynard} describe how local imitation-based interactions within a population aggregate into a \emph{single} RL entity at the macroscopic level.
Nevertheless, one may picture more complex scenarios where multiple populations would interact \ksr{(see \cref{fig:marl})}.
Our theory can be naturally extended to such multi-population scenarios:
\begin{itemize}
    \item All individuals from each population would evaluate their current strategy by playing a general $m$-player game with individuals sampled from the \emph{other} populations,
    \item Subsequently, they would execute an imitation step within their \emph{own} population.
\end{itemize}
\begin{figure}[tbp]
  \centering
  \scalebox{0.65}{\def\svgwidth{\linewidth}
\begingroup%
  \makeatletter%
  \providecommand\color[2][]{%
    \errmessage{(Inkscape) Color is used for the text in Inkscape, but the package 'color.sty' is not loaded}%
    \renewcommand\color[2][]{}%
  }%
  \providecommand\transparent[1]{%
    \errmessage{(Inkscape) Transparency is used (non-zero) for the text in Inkscape, but the package 'transparent.sty' is not loaded}%
    \renewcommand\transparent[1]{}%
  }%
  \providecommand\rotatebox[2]{#2}%
  \newcommand*\fsize{\dimexpr\f@size pt\relax}%
  \newcommand*\lineheight[1]{\fontsize{\fsize}{#1\fsize}\selectfont}%
  \ifx\svgwidth\undefined%
    \setlength{\unitlength}{103.00349808bp}%
    \ifx\svgscale\undefined%
      \relax%
    \else%
      \setlength{\unitlength}{\unitlength * \real{\svgscale}}%
    \fi%
  \else%
    \setlength{\unitlength}{\svgwidth}%
  \fi%
  \global\let\svgwidth\undefined%
  \global\let\svgscale\undefined%
  \makeatother%
  \begin{picture}(1,0.52046779)%
    \lineheight{1}%
    \setlength\tabcolsep{0pt}%
    \put(0,0){\includegraphics[width=\unitlength,page=1]{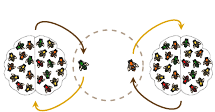}}%
    \put(0.85511239,0.01177246){\color[rgb]{0,0,0}\makebox(0,0)[lt]{\lineheight{1.25}\smash{\begin{tabular}[t]{l}$\mathcal{P}_2$\end{tabular}}}}%
    \put(0.09575624,0.00914385){\color[rgb]{0,0,0}\makebox(0,0)[lt]{\lineheight{1.25}\smash{\begin{tabular}[t]{l}$\mathcal{P}_1$\end{tabular}}}}%
    \put(0,0){\includegraphics[width=\unitlength,page=2]{multi_hive_marl_manual.pdf}}%
    \put(0.16335365,0.44282111){\color[rgb]{0,0,0}\makebox(0,0)[lt]{\lineheight{1.25}\smash{\begin{tabular}[t]{l}Strategy evaluation step\end{tabular}}}}%
    \put(0.68279949,0.44263099){\color[rgb]{0,0,0}\makebox(0,0)[lt]{\lineheight{1.25}\smash{\begin{tabular}[t]{l}Imitation step\end{tabular}}}}%
  \end{picture}%
\endgroup%
}
  \caption{Multi-population interaction.}
\label{fig:marl}
\end{figure}
When the imitation model is imitation of success (resp. weighted voter),
Proposition~\ref{prop:taylor} (resp. \ref{prop:maynard}) implies that such multi-population interactions can be modeled by a single CL (resp. MCL) agent per population.
The replicator dynamic remains the theoretical
bridge between the aggregate effect of local imitation within each population and its corresponding macro RL
agent, but becomes coupled with the replicator dynamic of other populations~\cite{Sato03} \ksr{(see \ref{app:multipopulation} for more details)}.
These macro RL agents then interact with each other as in a standard Multi-Agent Reinforcement Learning (MARL) scenario~\cite{tan1993multi, foerster2018deep, zhang2021multi}. Each of these agents optimizes its own rewards in a non-stationary environment, where rewards change as a function of the policies of all agents.
Note that this setting has similarities with environmental non-stationarity in single-population scenarios.
}

\subsection{\yb{Multi-level reinforcement learning}}
\yb{
Because our work focuses on imitative behaviors, both $R_\text{success}$ and $R_\text{wvoter}$ are revision protocols modeling pure imitation.
These protocols naturally aggregate into variants of the Replicator equation, which in turn correspond to coherent RL rules of the macro-organism.
However, in more complex scenarios, individuals themselves can change their type via
individual-level
RL in games against nature~\cite{biesmeijer2005use, Hunt20, maboudi20} or
in general population games~\cite{bouteiller2026sociodynamics}.
Without imitation,
deriving macro-agent rules for RL revision protocols
in games against nature
is trivial because there are no interactions between individuals in such revision protocols: the macroscopic RL agent is directly the distribution of types yielded over time by the $N$ parallel individual RL revision protocols, without any social influence~\cite{Hunt20}.
However, even without imitation, deriving analytical macro-agent rules for RL revision protocols in general population games is harder~\cite{bouteiller2026sociodynamics}.
In both cases, our framework allows adding an imitation step as long as the decoupling assumption of \cref{sec:modeling} is respected (i.e., individuals must resample an opponent to imitate after the evaluation step).
We leave the derivation of this type of hybrid dynamics for future work.\\
\newline
The population-policy equivalence abstracts individual-level policies $k$ as types, and enumerates all possible individual-level policies as the ``arms'' of the swarm-level policy $\pi$.
Note that the individual types $k \sim \pi$ describe all possible individual policies (which can be, e.g., stateful episodic policies), whereas the ``hive mind'' policy $\pi$ is always a stateless multi-armed bandit policy.
}

\subsection{\yb{Limitations and future work}} \label{sec:limitation}
\subsubsection{\yb{Biology}} \label{sec:limitation_bio}
\yb{
In this paper, we have shown that the swarm intelligence corresponding to the weighted voter model is MCL.
From a biologist perspective, it is however important to stress that the weighted voter model is a very partial description of actual honey bee behavior during nest-site selection, and thus MCL is not an exhaustive description of this biological process.
The weighted voter model is a discrete-time approximation of the waggle dance through which scout bees committed to a site candidate recruit uncommitted bees before returning to an uncommitted state themselves.
Several competing processes have been observed by biologists in real honey bee swarms~\cite{Seeley12} and are neglected in this model.
Therefore, a clear avenue for biology is to derive similar swarm-level RL rules from more accurate local interaction models of honey bees~\cite{marshall09, Reina_2017}, taking into account these other processes~\cite{Seeley12}:
\begin{itemize}
    \item \textbf{Spontaneous commitment:} Uncommitted bees occasionally commit to random options without encountering a waggle dance. From a macro-agent perspective, this mutation process may have an effect similar to $\varepsilon$-greedy exploration~\cite{sutton2018reinforcement}. Because after a while committed scouts revert to an uncommitted state, spontaneous commitment theoretically creates deadlocks around options of similar qualities.
    \item \textbf{Stop signals:} Committed scouts can produce stop signals to actively stop other committed scouts from broadcasting, targeting either their own type or other types depending on the collective decision phase (i.e., depending on $\pi$). These stop signals have been shown to resolve the aforementioned theoretical deadlocks.
    \item \textbf{Direct commitment switches:} Scouts may get recruited before they finish advertising, reducing their effective sampled quality estimates in the weighted voter model. This effect can be modeled by a dependence on $\pi$ of $r(k, \pi)$.
    \item \textbf{Quorum sensing:} The nest-site selection process of honey bees terminates early, when a quorum is reached. This process trivially corresponds to early-switching to a deterministic policy $\pi$ when the corresponding condition is met.
\end{itemize}
}

\subsubsection{\yb{Population structure}}
\ks{
Important factors from real populations
that were not modeled in our framework are: 
\begin{itemize}
    \item \textbf{Spatial structure:} The well-mixed assumption of \cref{prop:maynard}, may not hold in the presence of spatial structure. For instance, assortment among similar types of individuals in the population can lead to biased interactions and impact the decision-making process.
    \item \textbf{Heterogeneity:} Our model assumes homogeneous populations. However, populations can be heterogeneous. For instance, through the presence of
    \ybr{heterogeneous characteristics (age, size, capabilities, etc.)~\cite{beshers2001models}}
    or individuals following \ybr{their own} local update rules.
\end{itemize}
We leave the investigation of spatial structure and heterogeneity in our framework for future work.}

\subsubsection{\yb{Exploration-exploitation dilemma}}
\yb{
As seen in \cref{fig:combined}, finite-size effects may prevent CL and MCL to converge to the optimal strategy in finite populations.
This is because pure imitation can lead to situations in which a given type ceases to exist entirely in a finite population.
This issue becomes even more critical in real-world environments, which are often non-stationary~\ybr{\cite{schmickl2004costs, Prasetyo2019}}.
For instance, the quality of a food source may degrade over time in a foraging task.
With enough parallel individuals, the pure imitation protocols studied in this paper converge to the optimal strategy in stationary environments, but they don't provide a mechanism for escaping these strategies later on in non-stationary environments.
Thus, \ybr{real swarms need other mechanisms to continuously explore the environment in the presence of non-stationarity}.
For instance,
\ybr{ant colonies have been observed to maintain persistent stochasticity in their allocation of effort across options, which has been modeled as probability matching}~\cite{Hunt20}. \ksr{It has also been suggested that heterogeneity in ant colonies with the presence of ``wandering workers" (i.e., that do not recruit other ants but instead only forage for better alternatives) may help with exploration at the colony level~\cite{valentini2020}.} 
In the case of nest-site selection by honey bees, a similar effect is achieved via spontaneous commitment \ybr{to random options}~\cite{Seeley12}.
}


\subsection{\ks{Related work}} \label{rel:CDM}

\ks{
\textbf{Reinforcement Learning and Replicator Dynamic.}
The connection between RL and the TRD was first formulated by
B{\"o}rgers and Sarin~\cite{borgers1997learning}, who showed that two independent Cross Learning agents playing a 2-player stateless game follow a coupled Replicator Dynamic. Sato and Crutchfield~\cite{Sato03}, and Tuyls et al.~\cite{tuyls03b}, have independently demonstrated the same connection for a system of multiple Q-learning agents.
Hennes et al.~\cite{hennes09} introduced the state-coupled replicator dynamic,
extending this connection to stateful scenarios.
This connection
has proven useful in deriving new reinforcement learning algorithms such as Extended Cross Learning~\cite{tuyls03a} and Neural Replicator Dynamics~\cite{hennes20}.
We make a similar connection between the MRD and MCL.
}\\

\noindent
\ks
{
\textbf{Reinforcement Learning in insect colonies.}
Multi-armed bandits (MABs) naturally describe situations where one seeks the best choice out of $n$ options.
This is typically the case of Collective Decision Making in Swarm Intelligence.
Hunt et al.~\cite{Hunt20, hunt2020bayesian, baddeley2019optimal}, argue that ant colonies can be modeled as a ``superorganism'' performing a process analogous to Approximate Bayesian Computation.
Interestingly, their work further hypothesizes that individual ants themselves are Thompson Sampling agents rather than pure imitators.
During nest-site selection, bees are primarily imitators-inhibitors~\cite{Seeley12}, but individual reinforcement has been observed in their foraging behavior~\cite{biesmeijer2005use, maboudi20}.
While focusing on pure imitation allows us to make a clear point that reinforcement can emerge primarily at the swarm level, extending our derivations to multi-level reinforcement is an avenue for future work.
}\\

\noindent
\ks
{
\textbf{Collective Decision-Making in Swarm Robotics.}
Inspired by nest-site selection in honey bees, researchers have also developed collective decision-making strategies for swarm robotics~\cite{heikobook18}.
A representative setup~\cite{Soma2023} consists of
several
spatially separated sampling zones and a central nest zone, where robots must collectively identify the highest quality zone.
In the imitation step, robots update their opinions via various imitation protocols such as the weighted voter model,
the majority rule~\cite{valentini2016collective},
and cross-inhibition~\cite{Reina_2017}. 
Further studies have explored dynamic~\cite{Prasetyo2019} and multi-feature~\cite{Ebert2018} option qualities, continuous-space         
options~\cite{Raoufi21}, Bayesian belief modeling~\cite{Ebert2020}, and quality magnitude sensitivity~\cite{Pirrone2022}.
These works also provide a testbed for examining the influence of finite-size effects, congestion~\cite{Soma2023}, and connectivity constraints.
}

\subsection{Broader significance}

\begin{itemize}
\item
\yb{\textbf{Biology.}} 
\yb{Propositions \ref{prop:taylor} and \ref{prop:maynard} formally illustrate how pure imitation aggregates into a form of collective intelligence, which is that of a ``hive mind'' learning through reinforcement.
In particular, \cref{prop:maynard} provides partial theoretical grounding to previous observations of similar population-level agency in biological swarms of honey bees~\cite{Passino2008}, by explicitly mapping the waggle dance to swarm-level RL in the form of massively parallel MCL. Note that MCL is however incomplete as a model of bee swarm agency that would fit empirical observation, as it ignores spontaneous commitment, direct switches, cross-inhibition and quorum-sensing~\cite{Seeley12}.
Similar swarm-level agency has also extensively been observed in other social organisms~\cite{detrain1999information, baddeley2019optimal, Hunt20, hunt2020bayesian}, where our work can help build similar bridges between local imitation and global agency.}
In fact, parallel findings on pheromone-guided C. elegans build on the framework established in our work~\cite{vellinger2025pherorl}, suggesting that stigmergic coordination may also be \yb{understood} as a form of collective reinforcement learning.
Our analysis suggests that \yb{the waggle dance} is an example of reinforcement strategy having evolved at the group level, agreeing with the commonly accepted theory of kin selection to explain the behavior of honey bees~\cite{robinson89}.
\item
\textbf{Social dynamics and economy.}
Evolutionary Game Theory goes beyond modeling biology: it extends to many types of population dynamics, most notably in economics~\cite{borgers1997learning, apesteguia2007imitation}.
When thinking of $R_{\text{success}}$ as a model of certain decision-making strategies, \cref{lem:taylor} yields that 
\yb{local imitation aggregates into}
a macroscopic RL process that \ks{substantially} benefits the whole group, by mutualizing speed of convergence to optimal strategies (see \cref{sec:lr_batching}).
This opens an avenue for modeling the group-level impacts of economic actors' imitative behaviors as forms of collective reinforcement learning.
\item 
\yb{\textbf{Swarm robotics and swarm-inspired algorithms.}}
\ybr{These fields} take inspiration from complex emergent behaviors arising from a collective of natural entities following simple, local, and decentralized rules~\cite{swarmintelligence99}, to engineer algorithms that leverage collective principles like coordination, cooperation, and communication to tackle various problems (including CDM) across multiple domains~\cite{swarmdorigo2021,heikobook18,dorigoACO06}.
\cref{lem:taylor} and \cref{prop:maynard} provide theoretical grounding for the phenomenon commonly referred to as ``emergence" in these fields~\cite{de2004emergence}.
This bridge also opens up the possibility of porting ideas from Reinforcement Learning to derive \ks{local imitation-based} Swarm Intelligence rules. 
\end{itemize}

\subsection{Conclusion}

We have demonstrated that imitation-based collective behavior in large populations can be mathematically equivalent to reinforcement learning, providing a unifying framework that reinterprets swarm intelligence, social dynamics, and evolutionary processes as emergent forms of collective learning. This result provides a formal basis for understanding how simple, local imitation behaviors, often regarded as non-cognitive, can give rise to group-level intelligence and adaptive learning at scale. In particular, \cref{prop:maynard}
\ybr{implies}
that the ``hive mind''
observed in swarms of
bees can indeed be viewed as a single coherent agent learning via reinforcement, through a bandit algorithm that we coin \textit{Maynard-Cross Learning}.




\section*{Declarations}
\bmhead{Competing Interests}
Heiko Hamann holds the position of Editor-in-Chief of Swarm intelligence. To avoid any potential conflict of interest, they were completely excluded from the peer review, handling, and final decision-making processes for this submission.

\bmhead{Code availability} 
The code used for simulations is open-source and available at \url{https://github.com/MISTLab/HiveMindRL.git}.

\bmhead{Funding}
This work was funded by NSERC Discovery Grant No: 2019-05165.


\begin{appendices}
\section{Proofs for \cref{sec:theory}}\label{sec:apx_proof}
\PG*
\begin{proof}
Let $P(a \leftarrow b)$ denote the inflow of individuals of type $b$ into type $a$, i.e, the proportion of the population leaving type $b$ and adopting type $a$.
The population has a proportion of $\pi_b$ individuals of type $b$, each having a probability $\pi_a$ of meeting an individual of type $a$, and a conditional probability $\mathbb{E}[r_a]$ of switching to its type. Thus, we get $P(a\leftarrow b) = \pi_b\pi_a\mathbb{E}[r_a]$:

\begin{align} 
d\pi_a &= \sum_{b \neq a} \underbrace{P(a\leftarrow b)}_{\text{inflow}} - \underbrace{P(b\leftarrow a)}_{\text{outflow}} \\
           &= \sum_{b \neq a} \pi_b\pi_a\mathbb{E}[r_a] - \pi_a\pi_b\mathbb{E}[r_b]\notag\\
            &= \pi_a\Bigr[\sum_{b \neq a} \pi_b\mathbb{E}[r_a] - \sum_{b \neq a}\pi_b\mathbb{E}[r_b]\Bigr]               \hspace{2em}  \sum_{b \neq a} \pi_b + \pi_a = 1\notag\\
			  &= \pi_a\Bigr[(1-\pi_a)\mathbb{E}[r_a] - \sum_{b \neq a}\pi_b\mathbb{E}[r_b]  \Bigr]  \notag\\
			  &= \pi_a\Bigr[\mathbb{E}[r_a] - \sum_{b}\pi_b\mathbb{E}[r_b]\Bigr] \notag\\
            &=\pi_a (q^\pi_a - v^\pi)
\end{align}
\end{proof}

\RL*
\begin{proof}
Let us compute the expectation over actions sampled from $\pi$ in Eq.~\ref{eq:ecl}.
For convenience, we write\\ 
$\mathbb{E}[d\pi_a] := \mathbb{E}_{k \sim \pi}[ d\pi_a(k, \yb{\pi})]$, and
$\mathbb{E}[r_k]:=\mathbb{E}_{r_k \sim r(k, \yb{\pi})}[r_k]$:
\begin{align}
\mathbb{E}[d\pi_a] &= \sum_{k=1}^n \pi_k . d\pi_a(k) \\
   &= \pi_a . d\pi_a(a) + \sum_{k \neq a} \pi_k . d\pi_a(k) \notag\\
    &= \pi_a \mathbb{E}[r_a](1 - \pi_a) + \sum_{k \neq a} \pi_k \mathbb{E}[r_k](- \pi_a) \label{eq:lem1_ref1} \notag\\
    &= \pi_a\Bigr[\mathbb{E}[r_a] - \pi_a \mathbb{E}[r_a] - \sum_{k \neq a} \pi_k \mathbb{E}[r_k]\Bigr]\notag\\
    &= \pi_a\Bigr[\mathbb{E}[r_a] - \sum_{k} \pi_k \mathbb{E} [r_k]\Bigr]\notag\\
    &= \pi_a (q^\pi_a - v^\pi)
\end{align}
\end{proof}



\section{Simulations}\label{sup:methods}
In this section, we provide some additional results to consolidate the theory presented in the manuscript.
To do so, we first describe the two RL update rules: CL and MCL, in both their streaming and parallel variants. We also outline the implementation details of the population update rules $R_{\text{success}}$ and $R_{\text{wvoter}}$ along with their variants. Finally, we numerically simulate the TRD and MRD to compare the RL and population update rules against their corresponding analytical solutions. It is important to note that MCL is not intended as a competitive bandit algorithm, but only as a biologically motivated construct designed to demonstrate that an imitation-based population update ($R_{\text{wvoter}}$) can be captured by an RL update rule.

\subsection{Environment}
We consider the standard multi-armed stateless bandit setting described in preliminaries (see \cref{prelim:cl}).
It is clear from the Population-policy equivalence remark (\cref{re:equivalence}) that we can use the same environment for RL and population experiments. \ks{In these experiments, we only consider stationary reward distribution}, where the environment returns a noisy reward signal sampled from the hidden distribution~$r(a)$ when action $a$ is taken. \ks{Note that the reward distribution is independent of $\pi$.}
We define the hidden reward distribution as a uniform distribution, given by $r \sim \mathcal{U}(\ks{q_{a}} - \Delta, \ks{q_{a}} + \Delta)$, where $\ks{q_{a}}$ is the mean reward associated with action $a$, and $\Delta$ controls the amplitude of noise around $\ks{q_{a}}$. To ensure the validity of both RL and population updates, rewards are bounded within the interval [0,1], which imposes the constraint $\ks{q_{a}} - \Delta \geq 0$ and $\ks{q_{a}} + \Delta \leq 1$. We consider three distinct scenarios for the reward means across actions: (i) Low, where all $\ks{q_{a}}$ values are equally spaced in the range [0.1, 0.4], (ii) Middle, where $\ks{q_{a}}$ values are equally spaced in the range [0.4,0.7], and (iii) High, where all $\ks{q_{a}}$ values are equally spaced in the range [0.6,0.9]. \ks{For experiments with a non-stationary reward distribution see~\cref{app:congestion}}.

\subsection{RL Experiments}
\textbf{Streaming:} In these experiments, an RL agent interacts with a single environment in a streaming fashion. The RL agent starts with an initial random policy $\pi$. The agent then samples one action~$k$ at each computation step from $\pi$ in an iterative fashion.  For pulling this action~$k$, the agent receives a noisy reward signal~$r_k \sim r(k)$ from the environment. Subsequently, for CL, the agent \ks{uses} \cref{eq:cl} with learning rate $\alpha$ to update the policy.  
\begin{equation}
	\forall a, \pi_a \leftarrow \pi_a +  \alpha r_k\begin{cases}
		1 - \pi_a &  \text{if}\ a=k\\
		- \pi_a &  \text{otherwise}
	\end{cases}\;\;
\end{equation}
Whereas for MCL, \cref{eq:mcl_alpha} cannot be used directly, since $v^{\pi}$ is not estimated. Therefore,  $v^{\pi}$ is approximated by employing a moving average over rewards, where $\gamma$ is a weighting factor for recent rewards:
\begin{equation}
    \bar{r} \leftarrow \gamma r + (1-\gamma) \bar{r}\;.
\end{equation}
Moreover, since this update rule can make $\pi$ invalid, i.e., components could become negative or above one, we clamp $\pi$ between 0 and~1: 
\begin{equation}
	\forall a, \pi_a \leftarrow  \text{clamp}\left(\pi_a +  \alpha\frac{r_k}{\bar{r}}\begin{cases}
		1 - \pi_a &  \text{if}\ a=k\\
		- \pi_a &  \text{otherwise}
	\end{cases}\;\;\right)
\end{equation}
Where $\alpha$ is the learning rate. 
These computations are carried out for every training \emph{step}, and there are $S$ steps per iteration.\\
\textbf{Parallel:} In these experiments, we implement parallel variants of the CL and MCL update rules, referred to as P-CL and P-MCL hereafter, respectively. P-CL performs a straightforward parallelization of the CL rule: at each update step, the policy $\pi$ is updated based on the average effect of $B$ independent samples collected from $B$ parallel environments, as if simulating the CL update $B$ times in parallel and averaging the results. Similarly, in P-MCL, the policy update is computed using the average effect of $B$ parallel samples, normalized by $v^{\pi}$. Unlike streaming MCL, where normalization ($v^{\pi}$) is based on a moving average of past rewards, P-MCL uses the current mean of the $B$ parallel rewards as the normalizing factor. With P-MCL, we also need to explicitly limit these policy updates between 0 and 1 to ensure that $\pi$ remains valid. These calculations are also performed for $S$ training steps per iteration.

\subsection{Population Experiments}\label{app:pop_details}
\textbf{$R_{\text{success}}$}:
We implement $R_{\text{success}}$ from \cref{prelim:success}. We start with an equal proportion of individuals associated with any type. Further, each individual receives a stochastic payoff estimate ($r \sim r(a)$) for their type. Then, at each step, everyone is paired with another random individual for imitation. All individuals then generate a random real number between 0 and 1, and if the random real number is greater than the payoff of their paired individual, imitation is successful and they switch to their paired partner's type (rule~3 of $R_{\text{success}}$). If the generated random number is not greater than the payoff of their paired individual, they do not imitate and stick to their own types. Further, we consider another variant \textit{Deterministic Imitation of Success} where individuals deterministically switch to the imitating partner's type if the rewards of the partner are higher than their rewards. There are $S$ decision-making steps per iteration.\\
\textbf{$R_{\text{wvoter}}$}:
We implement $R_{\text{wvoter}}$ from \cref{prelim:wvoter}. We start with an equal proportion of individuals associated with any opinion. Further, each individual receives a stochastic quality estimate ($r \sim r(a)$) for their opinion. Then at each step, each individual ($i$) switches to a \ks{option} sampled from the distribution of votes cast $v^{(i)}$ (simulated effect of ``waggle dance"), where $v_k^{(i)}$ is the ratio of votes cast for option $k$ by the total number of votes cast in \ks{its} neighborhood ($\mathcal{N}^{i}$) excluding itself: 
\begin{equation}
    v_k^{(i)} = \frac{\sum\limits_{\forall p \in \mathcal{N}^{i} :\ \ks{T_p = k}} r_p}{\sum\limits_{\forall q \in \mathcal{N}^{i}} r_q}\;.
\end{equation}
These neighborhoods for each individual are formed by randomly sampling $M$ individuals from $\mathcal{P}$ at every time step. It is to be noted that, unless specified, the neighborhood sizes are equal to the population size. Further, we consider two variants \textit{Deterministic  Imitation of Success with Weighted Voter rules} and \textit{Stochastic Imitation of Success with Weighted Voter rules}, where we combine ideas from $R_{\text{wvoter}}$ and $R_{\text{success}}$. With both of these variants, each individual is first paired with one of their neighbors with a probability proportional to the neighbor's rewards. After this pairing, with Stochastic Imitation of Success with Weighted Voter Rules, individuals switch similar\ks{ly} to $R_{\text{success}}$ and with Deterministic Imitation of Success with Weighted Voter rules, individuals switch similar\yb{ly} to Deterministic Imitation of Success. There are $S$ decision-making steps per iteration. The code for reproducing the simulations can be found here \url{https://github.com/MISTLab/HiveMindRL.git}\\

\noindent
\ks{All the aforementioned population and RL update rules can be easily extended to the non-stationary setting, as the difference between the stationary and non-stationary setting is only in the reward generation process (or strategy evaluation step). The imitation rules of individuals and RL update rules of agents are indifferent to the reward or payoff generation process.}

\subsection{TRD and MRD} \label{app:trd_mrd}
To empirically validate \cref{lem:taylor} and \cref{prop:maynard}, we numerically simulate both the variants of RD according to \cref{eq:trd,eq:mrd}. As these equations are continuous, we discretize them by a step $\delta$ (discretizing step). Further, we start from an initial random population/policy ($\pi$) and simulate its evolution according to TRD and MRD between time intervals $[0,t_f]$, using the privileged information $\ks{q_a}$ not available to RL and population experiments.  
\begin{align*}
      \pi_a \leftarrow \pi_a + \delta \pi_a [\ks{q_a} - \sum_l \pi_l \ks{q_l}]  
      &&                                      
      \pi_a \leftarrow \pi_a + \delta \frac{\pi_a}{v^{\pi}} [\ks{q_a} - \sum_l  
  \pi_l \ks{q_l}]                                                             
\end{align*}  
\ks{For non-stationary settings such as in \cref{app:congestion}, $q_a$'s are replaced by $q^{\pi}_a$'s.}
\ks{\begin{align*}
      \pi_a \leftarrow \pi_a + \delta \pi_a [q^{\pi}_a - \sum_l \pi_l q^{\pi}_l]  
      &&                                      
      \pi_a \leftarrow \pi_a + \delta \frac{\pi_a}{v^{\pi}} [q^{\pi}_a - \sum_l  
  \pi_l q^{\pi}_l]                                                             
\end{align*}  }

\begin{table}[htb]
    \centering
    \begin{subtable}{.33\linewidth}
    \begin{tabular}{|c|c|}
         \hline
         Hyperparameter & value \\
         \hline
         Arms ($n$) & 10  \\
         iterations & 1000 \\
         Range of noise ($2\Delta)$ & 0.2\\
         Parallel Environments ($B$) & $\{10, 1000 \}$ \\
         Discretizing factor ($\delta$) & 1 \\ 
         \hline
    \end{tabular}
    \caption{Parallel RL experiments}
    \end{subtable}
    \begin{subtable}{.33\linewidth}
        \begin{tabular}{|c|c|}
         \hline
         Hyperparameter & value \\
         \hline
         Types/options ($n$) & 10  \\
         Seeds & 1000 \\
         Range of noise $(2\Delta)$ & 0.2\\
         Population size ($N$) & $\{10, 1000 \}$ \\
         Discretizing factor ($\delta$) & 1 \\ 
         \hline
    \end{tabular}
    \caption{Population experiments}
    \end{subtable}
        \begin{subtable}{.33\linewidth}
    \begin{tabular}{|c|c|}
         \hline
         Hyperparameter & value \\
         \hline
         Arms ($n$) & 10  \\
         Learning rate  ($\alpha$) & $\{0.001, 0.1\}$ \\
         iterations  & 1000 \\
         Range of noise ($2\Delta)$ & 0.2\\
         Weight factor ($\gamma$) & 0.01 \\ 
         Discretizing factor ($\delta$) & $\alpha$ \\ 
         \hline
    \end{tabular}
    \caption{Streaming RL experiments}
    \end{subtable}
            \begin{subtable}{.33\linewidth}
    \begin{tabular}{|c|c|}
         \hline
         Hyperparameter & value \\
         \hline
         Types/options ($n$) & 10  \\
         Population size & 1000 \\
         iterations  & 1000 \\
         Range of noise ($2\Delta)$ & 0.2\\
         Neighborhood sizes & $\{2, 10, 1000\}$ \\ 
         Discretizing factor ($\delta$) & $\alpha$ \\ 
         \hline
    \end{tabular}
    \caption{Neighborhood experiments}
    \end{subtable}
   \caption{Hyperparameters for RL and population experiments.}
    \label{sup:hyper}
\end{table}

In all the subsequent experiments, we track the evolution of optimal decisions in both reinforcement learning and population experiments. In the RL setting, we monitor the ``$\%$ optimal action", defined as $\pi_a \times 100$ (where $a$ is the optimal action), and report the average along with its standard deviation over all seeds at each step. Similarly, in the population setting, we compute the ``$\%$ optimal type", defined analogously as $\pi_a \times 100$, where $a$ is the optimal type. We also include the trajectories of the optimal action/type under the TRD and MRD dynamics to compare the empirical behavior of the RL and population update rules with their corresponding analytical models. The hyperparameters used in these simulations are provided in \cref{sup:hyper}.\\
\subsection{Streaming RL update rules follow analytical solutions when the learning rate is small} \label{app:streaming_rl}
\begin{figure*}[h]
    \centering
    \includegraphics[width=\linewidth]{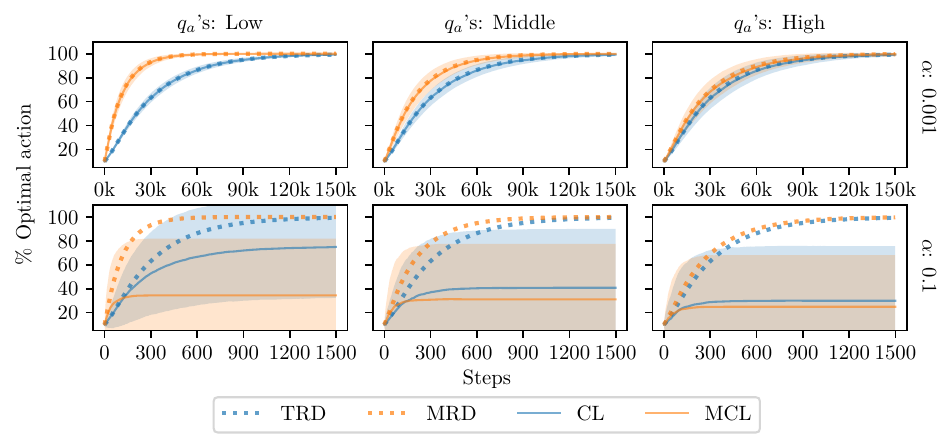}
    \caption{Results for streaming RL experiments.}
    \label{fig:stream_rl}
\end{figure*}
These results are presented in \cref{fig:stream_rl}.
For all scenarios, CL and MCL follow TRD and MRD, respectively, with small $\alpha$, which can be explicitly seen with the dotted line of the analytical solutions (TRD, MRD) exactly at the center of the optimal $\%$ action curves of the CL and MCL update rules. This empirically validates that, with a small $\alpha$, \cref{eq:cl} with $\alpha$ and 
\cref{eq:mcl_alpha} follow the TRD and MRD, respectively, even in a streaming fashion.
However, as soon as $\alpha$ increases, CL and MCL start deviating from their respective analytical solutions and have a huge standard deviation.
This is a well-known effect in optimization literature.  Interestingly, imagining the action samples forming a population (see \cref {sec:lr_batching}), we see that a larger $\alpha$ corresponds to a smaller population, which leads to a poor approximation of the expected update. \\

\subsection{Parallel RL update rules follow analytical solutions when the number of parallel environments is large.} \label{app:parallel_rl}
\begin{figure}[htbp]
    \centering
    \includegraphics[width=\linewidth]{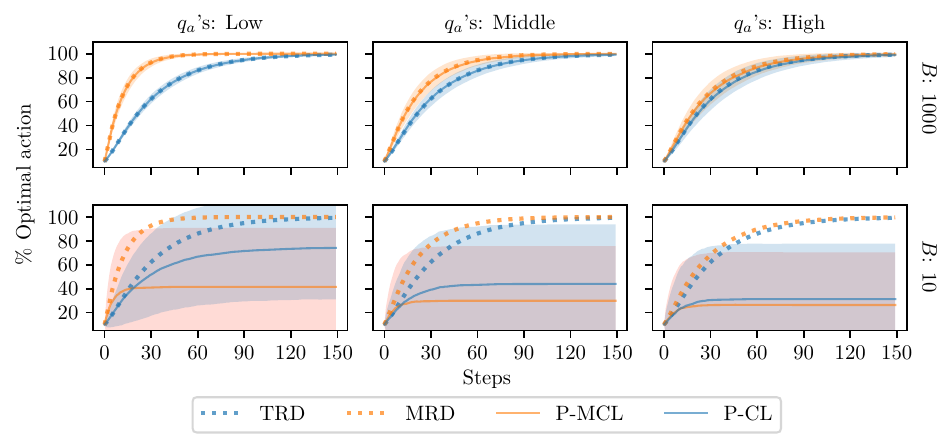}
    \caption{Results for parallel RL experiments. }
    \label{fig:prallel_rl}
\end{figure}
As seen in \cref{fig:prallel_rl}, it is clear that P-CL and P-MCL follow TRD and MRD, respectively, when updates are made utilizing a large number of parallel environments (this can be seen from the way the analytical solution is exactly at the center of the $\%$ optimal action curves of P-CL and P-MCL).
However, as soon as the number of parallel environments is reduced, the updates deviate from their analytical solutions (see \cref{sec:lr_batching}).\\

\subsection{Population update rules: $R_{\text{success}}$ \& $R_{\text{wvoter}}$ follow their analytical solutions when the population sizes are large} \label{app:population}
\begin{figure*}[h]
    \centering
    \includegraphics[width=\linewidth]{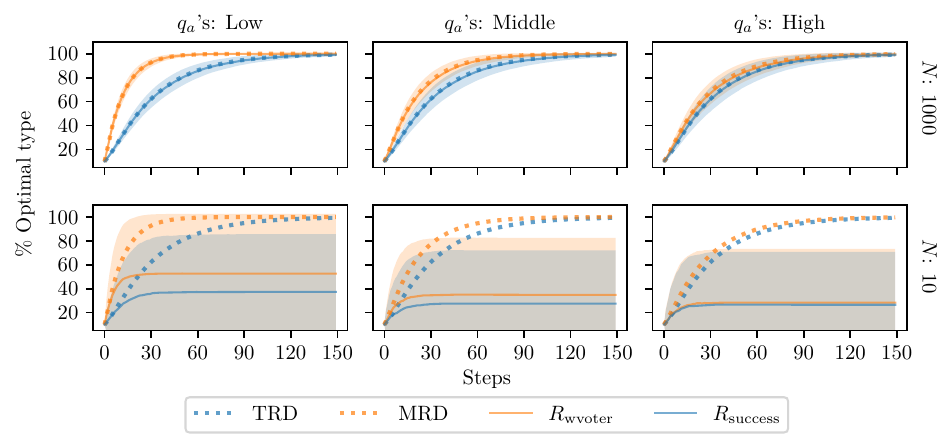}
    \caption{Results for population experiments. }
    \label{fig:population}
\end{figure*}
It can be seen in \cref{fig:population} that both $R_{\text{success}}$ and $R_{\text{wvoter}}$ follow TRD and MRD respectively when the population size is large. As soon as the population shrinks, $R_{\text{success}}$ and $R_{\text{wvoter}}$ begin to deviate from the analytical solution. 

\subsection{$R_{\text{success}}$ \& $R_{\text{wvoter}}$ variants} \label{app:variants}
\begin{figure*}[h]
    \centering
    \includegraphics[width=\linewidth]{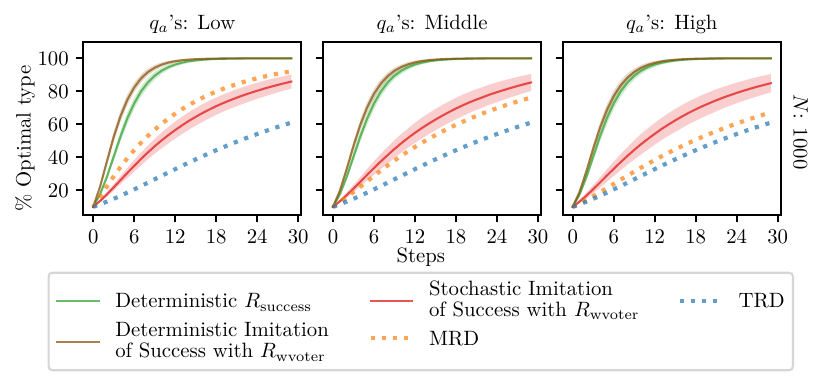}
    \caption{\ks{Results for variants of population update rules for $N$=1000.}}
    \label{fig:variants}
\end{figure*}
It can be seen in \cref{fig:variants} that the variants Deterministic Imitation of Success and Deterministic Imitation of Success with Weighted Voter rules perform better than MRD for all scenarios. However, Stochastic Imitation of Success with Weighted Voter Rules performs better than MRD in the Middle and High scenario, and performs worse than MRD in the low scenario. This could be because the probability of imitation might be low for the Low scenario as the scales of rewards are smaller.

\subsection{\ks{Neighborhood} sizes in $R_{\text{wvoter}}$ only affect the convergence speed to MRD} \label{app:neighoburhood}
\begin{figure*}[h]
    \centering
    \includegraphics[width=\linewidth]{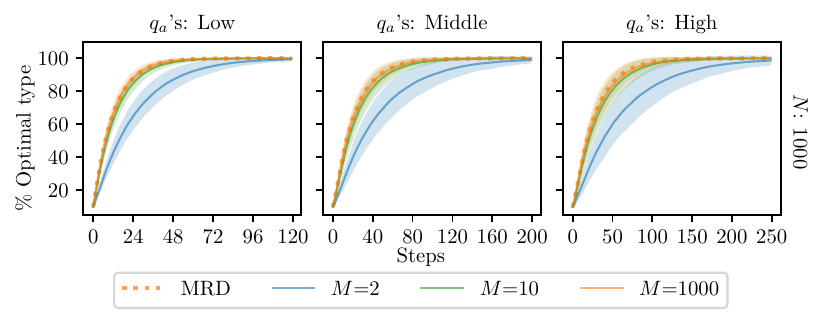}
    \caption{Results for neighborhood experiments.}
    \label{fig:neighborhood}
\end{figure*}
It can be seen in \cref{fig:neighborhood} that for a large enough population $R_{\text{wvoter}}$ follows MRD for any neighborhood size. However, it can be seen that the convergence speed is affected by smaller neighborhood sizes. \\

\subsection{Convergence rate of MRD is $\geq$ TRD} As noted in~\cite{sandholm2010population}, TRD and MRD can be rearranged in the form: 
\begin{align}
    \dot{\pi_a} &= v^{\pi}(\frac{\pi_a q^\pi_a}{v^{\pi}} - \pi_a) & \text{(TRD)} \\
   \dot{\pi_a} &=  1(\frac{\pi_a q^\pi_a}{v^{\pi}} - \pi_a)  & \text{(MRD)}
\end{align}
$\dot \pi$ being the update ``speed" and  $v^{\pi}$ being bounded between 0 and 1.
The MRD speed is thus greater than the TRD speed for a given scenario.
Empirically, we observe that MRD converges faster than TRD, especially when the $q_a$'s are low and middle, as seen with any of the \cref{fig:stream_rl,fig:prallel_rl,fig:population}.
Whereas, when the $q_a$'s are high, there is very little difference (as $v^{\pi} \approx 1$). By extension, this also implies that MCL (for small $\alpha$), P-MCL (for large $B$), and $R_{\text{wvoter}}$ (for large population) have convergence rates $\geq$ CL, P-CL, and $R_{\text{success}}$, respectively. However, to compare the convergence speeds of TRD and MRD across various reward scales, we revert the equations to their original form. 

\begin{align}
    \dot{\pi_a} &= \pi_a (q^\pi_a - v^{\pi}) & \text{(TRD)} \\
   \dot{\pi_a} &=  \pi_a \Big(\frac{q^\pi_a  - v^{\pi}}{v^{\pi}}\Big)  & \text{(MRD)}
\end{align}
\noindent
As the term $q^\pi_a - v^{\pi}$ denotes the relative fitness of any $a$ (or advantage in the RL literature), we can see that TRD has a constant convergence speed across the reward scales. However, with MRD, this relative fitness is normalized by $v^{\pi}$, which increases with higher reward scales, leading to slower convergence speed with higher reward scales.  This, combined with the observation that the speed of MRD $\geq$ TRD for any given scenario, shows that MRD catches up with TRD as the reward scales increase, as seen in \cref{fig:stream_rl,fig:prallel_rl,fig:population}.

\subsection{\yb{Decoupling imitation from evaluation}} \label{app:decoupling}
\yb{
When modeling pairwise interactions by 2-player games, $r_a \sim r(a, b)$ is a noisy estimate of the expected payoff of type $a$ against the population $\pi$, denoted $q^\pi_a$.
Of course, $r_a$ is also a less noisy estimate of the expected payoff of type $a$ against type $b$ (let us denote this quantity $q^b_a$), but our mathematical analysis does not hold if $i$ directly imitates $j$ based on $q^b_a$ (the inflow into type $a$ from type $b$ would become coupled to $q^b_a$ and $q^a_b$ instead of $q^\pi_a$ in the proof of \cref{lem:PG}).
This means that individuals need to randomly resample another partner to imitate for our derivations to hold, similar to \cite{weibull95, schlag1998imitate, sandholm2010population}.
While this decoupling may seem contrived when modeling local imitation, it holds when imitation happens at low frequency compared to quality estimation (for instance if quality estimates are an average over many interactions), or whenever quality estimation happens independently from imitation (which is true in the weighted voter model).
}

\subsection{\ks{Congestion experiments}} \label{app:congestion}
\ks
{In this experiment, we show empirical support for our theory in a non-stationary  environment. To motivate non-stationarity, consider the congestion example highlighted 
in~\cref{sec:policy_dependence}. In a foraging scenario, the quality of a site may reduce as a result of the number of individuals sampling that site. We model this in a 
simplified scenario involving a two-armed bandit where the rewards of both arms are:  
$r \sim \mathcal{U}(q - \Delta, q + \Delta) - \omega_a\pi_a$, which has a
linear congestion penalty defined by $\omega_a$. Note that $\omega_a \leq q -\Delta$ ensures $r \geq 0$, and $q + \Delta - \omega_a \leq 1$ ensures $r \leq 1$. Non-stationarity arises solely from the congestion penalty $\omega_a \pi_a$, as $q$ is policy independent. As a simple experiment, we set both arms to have equal quality $q = 0.7$ and equal half-width noise $\Delta = 0.1$, meaning that without the congestion penalty, the rewards for both arms are identical. We consider the congestion factors to be 0.1 for the arm with the lower congestion penalty (which is referred to as the site A) and 0.4 for the arm with the higher congestion penalty. Further, we set the number of individuals to be $2000$ in the population experiments and the number of parallel environments to $2000$ in the RL experiments and present the results over 1000 iterations in \cref{fig:congestion}. However, note that parameters we choose for this simulation are arbitrary, they are meant to be interpreted qualitatively not quantitatively.
} 

\subsection{\ksr{Coupled RD in multi-population interaction}}
\label{app:multipopulation}
\ksr
{
Our framework can be extended to multiple interacting populations. Game theory models such interactions via a payoff (game) matrix. Examples of such games include Prisoner's Dilemma, Stag Hunt, and Hawk Dove. Consider two interacting populations ($\mathcal{P}_1$ and $\mathcal{P}_2$), where individuals of $\mathcal{P}_1$ have strategies $i$ $\in$ $\{1,\dots,m\}$ and $\mathcal{P}_2$ have strategies $j$ $\in$ $\{1,\dots,n\}$. Let $A$ and $B \in \mathbb{R}^{m\times n}$ denote the payoff matrices that capture the interaction payoffs of $\mathcal{P}_1$ and $\mathcal{P}_2$. In the strategy evaluation step, when an individual of $\mathcal{P}_1$ playing strategy $i$ interacts with an individual of $\mathcal{P}_2$ playing strategy $j$, the payoff to $\mathcal{P}_1$'s individual is given by $A_{ij}$ and the payoff to $\mathcal{P}_2$'s individual by $B_{ij}$. From \cref{re:equivalence}, we know that both $\mathcal{P}_1$ and $\mathcal{P}_2$ have an associated population vector: $x$ and $y$. We consider these vectors in column form for this discussion. The action values for strategy $i$ of $\mathcal{P}_1$ and strategy $j$ of $\mathcal{P}_2$ are:
$$q^{x}_i := \sum_j A_{ij}y_j = (Ay)_i, \qquad q^{y}_j := \sum_i x_iB_{ij} =(x^\top B)_j$$  
The values of the population vectors can be written as:
$$v^{x} := \sum_i x_iq^{x}_i = x^\top Ay, \qquad v^{y} := \sum_j q^{y}_jy_j =
x^\top By$$
 As noted earlier, since the imitation rules of individuals are indifferent to the payoff generation process, our theory holds, meaning that a large population of individuals following $R_{\text{success}}$ (resp. $R_{\text{wvoter}}$) by imitating individuals within their population can be modeled as CL (resp. MCL) RL agent at the macro level. TRD and MRD remain the analytical solutions for the imitation and RL rules.
The only difference from the single population scenario is that the payoffs individuals in each population receive during the strategy evaluation step are a result of interaction with individuals from other populations, meaning that instead of $r(k,\pi)$, we now have $r^i(k,x,y)$ for $i \in \{1,2\}$.} \\

\noindent
\ksr{
Thus, in this multi-population scenario, TRD takes the form
\begin{align*}
      \dot{x_i} &= x_i \Big(q^{x}_i - v^{x}\Big) & \dot{y_j} &= y_j \Big(q^{y}_j
  - v^{y}\Big)
  \end{align*}
  \begin{align}
      \dot{x_i} &= x_i \Big((Ay)_i - x^\top Ay\Big) & \dot{y_j} &= y_j
  \Big((x^\top B)_j - x^\top By\Big)
  \end{align}
  Similarly, the MRD takes the form,
  \begin{align*}
      \dot{x_i} &= x_i \Big(\frac{q^{x}_i - v^{x}}{v^{x}}\Big) & \dot{y_j} &=
  y_j \Big(\frac{q^{y}_j - v^{y}}{v^{y}}\Big)
  \end{align*}
  \begin{align}
      \dot{x_i} &= x_i \Big(\frac{(Ay)_i - x^\top Ay}{x^\top Ay}\Big) &
  \dot{y_j} &= y_j \Big(\frac{(x^\top B)_j - x^\top By}{x^\top By}\Big)
  \end{align}
It is clear from the equations that the dynamics of each population depend on the other population's replicator dynamics, forming a coupled system. 
}

\end{appendices}

\bibliography{sn-bibliography.bib}

\end{document}